\documentstyle[12pt]{article}
\input epsf.tex

\newcommand{\bmat}{\left(\begin{array}}
\newcommand{\emat}{\end{array}\right)}

\def\preal{{\rm Re\,}}
\def\pim{{\rm Im\,}}

\def\yzero{\smash{\hbox{$y\kern-4pt\raise1pt\hbox{${}^\circ$}$}}}

\def\-{\hphantom{-}}
\def\ov{\overline}
\def\s2{\frac{1}{\sqrt2}}

\def\beq{\begin{equation}}
\def\eeq{\end{equation}}
\def\beqa{\begin{eqnarray}}
\def\eeqa{\end{eqnarray}}

\def\Tr{{\rm Tr \,}}
\def\diag{{\rm diag \,}}
\def\IR{\bf R}
\def\IC{\bf C}
\def\IT{\bf T}
\def\IZ{\bf Z}
\def\NN{{\cal{N}}}
\def\hi{\bf i}

\def\Dsl{\,\raise.15ex\hbox{/}\mkern-13.5mu D} %this one can be subscripted
\def\id{{\rm I}}
\catcode`\@=11
\newdimen\@rotdimen
\newbox\@rotbox

\def\@vspec#1{\special{ps:#1}}%  passes #1 verbatim to the output
\def\@rotstart#1{\@vspec{gsave currentpoint currentpoint translate
   #1 neg exch neg exch translate}}% #1 can be any origin-fixing transformation
\def\@rotfinish{\@vspec{currentpoint grestore moveto}}% gets back in synch
\def\@rotr#1{\@rotdimen=\ht#1\advance\@rotdimen by\dp#1%
   \hbox to\@rotdimen{\hskip\ht#1\vbox to\wd#1{\@rotstart{90 rotate}%
   \box#1\vss}\hss}\@rotfinish}
\def\@rotl#1{\@rotdimen=\ht#1\advance\@rotdimen by\dp#1%
   \hbox to\@rotdimen{\vbox to\wd#1{\vskip\wd#1\@rotstart{270 rotate}%
   \box#1\vss}\hss}\@rotfinish}%
\def\@rotu#1{\@rotdimen=\ht#1\advance\@rotdimen by\dp#1%
   \hbox to\wd#1{\hskip\wd#1\vbox to\@rotdimen{\vskip\@rotdimen
   \@rotstart{-1 dup scale}\box#1\vss}\hss}\@rotfinish}%
\def\@rotf#1{\hbox to\wd#1{\hskip\wd#1\@rotstart{-1 1 scale}%
   \box#1\hss}\@rotfinish}%
\def\rotate{\@ifnextchar[{\@rotate}{\@rotate[l]}}
\def\@rotate[#1]#2{\setbox\@rotbox=\hbox{#2}\@nameuse{@rot#1}\@rotbox}

\catcode`\@=12
\topmargin
-1.5cm
\textwidth
15.5cm
\textheight
23.5cm
\oddsidemargin
0.7cm
\evensidemargin
1.2cm

\begin{document}

%----------------------------------------------------------------------%
%  numbering equations with section number
%----------------------------------------------------------------------%
\makeatletter
\@addtoreset{equation}{section}
\makeatother
\renewcommand{\theequation}{\thesection.\arabic{equation}}
%----------------------------------------------------------------------%
%  title page
%----------------------------------------------------------------------%
\pagestyle{empty}
%\vspace*{1.0in}
\rightline{ IFT-UAM/CSIC-03-42}
\rightline{\tt hep-th/yymmnn}
\vspace{0.5cm}
\begin{center}
\LARGE{Flux-induced SUSY-breaking
soft terms  \\[10mm]}
\large{
P. G. C\'amara$^*$, L.~E.~Ib\'a\~nez$^{*,**}$ and A. M. Uranga$^*$
\\[2mm]}
\small{
*\ Departamento de F\'{\i}sica Te\'orica C-XI
and Instituto de F\'{\i}sica Te\'orica  C-XVI,\\[-0.3em]
Universidad Aut\'onoma de Madrid,
Cantoblanco, 28049 Madrid, Spain.\\ [4mm]
**\ Theory Division, CERN, 1211 Geneva 23, Switzerland.
\\[9mm]}
\small{\bf Abstract} \\[7mm]
\end{center}

\begin{center}
\begin{minipage}[h]{14.0cm}
{\small
We describe the computation of SUSY-breaking terms on a D3-brane in a quite 
general type IIB supergravity background. We apply it to study the 
SUSY-breaking induced on the D3-brane worldvolume by the presence of NSNS 
and RR 3-form fluxes. We provide explicit general formulae for the 
SUSY-breaking soft terms valid for the different types of fluxes, leading  
to different patterns of soft terms. Imaginary anti self-dual fluxes with 
$G_3$ a pure $(3,0)$-form lead to soft terms corresponding to 
dilaton-dominated SUSY-breaking. More general SUSY-breaking patterns are 
discussed, arising from more general fluxes, or from distant 
anti-D3-branes. The known finiteness properties of dilaton-dominated soft 
terms are understood in terms of holography. The above results are 
interpreted in the context of the 4d effective supergravity theory, where 
flux components correspond to auxiliary fields of e.g. the 4d dilaton and 
overall volume modulus. We present semirealistic Type IIB orientifold 
examples with (meta)stable vacua leading to non-vanishing soft terms of 
the dilaton-domination type. Such models have many of the ingredients of 
the recent construction of de Sitter vacua in string theory. We 
finally explore possible phenomenological applications of this form of 
SUSY-breaking, where we show that soft terms are of order $M_s^2/M_p$. Thus 
a string scale of order $M_s=10^{10}$ GeV, and compactification scale 
 three  orders of magnitude smaller could explain  the smallness of the 
weak scale versus the Planck mass.
}

\end{minipage}
\end{center}
\newpage
%----------------------------------------------------------------------%
%  Resetting of counters
%----------------------------------------------------------------------%
\setcounter{page}{1}
\pagestyle{plain}
\renewcommand{\thefootnote}{\arabic{footnote}}
\setcounter{footnote}{0}
%----------------------------------------------------------------------%
%  Paper begins
%----------------------------------------------------------------------%

\section{Introduction}

Low energy $\NN=1$ supersymmetry is   one of the most promising avenues 
to understand the stability of the electroweak scale against radiative 
corrections. Supersymmetry must however be broken, producing masses for 
the unobserved SUSY partners of quarks and leptons not much above the 
electroweak scale. One usually parametrizes supersymmetry breaking in terms 
of SUSY-breaking soft terms, i.e. ${\rm dim}<4$ operators which do not 
induce new quadratic divergences which would spoil the nice loop 
properties of $\NN=1$ supersymmetry. These soft terms result after 
spontaneous supersymmetry breaking, irrespective of what the specific 
source of SUSY-breaking may be.   

On the other hand, the structure of soft terms has a crucial impact on the
pattern of masses and couplings for the SUSY partners of the usual 
Standard Model (SM) particles, and hence it would be very important to 
have a theory which makes predictions about these quantities. If 
low-energy SUSY is discovered at LHC such a theory would be amenable to 
direct experimental test.

If string theory is indeed the fundamental theory underlying both the 
gravitational as well as the SM interactions of elementary particles, it 
should be able to describe the structure of soft terms. Indeed, attempts 
to understand the structure of soft terms resulting from SUSY-breaking 
in perturbative heterotic compactifications were done since the early days 
of heterotic string phenomenology \cite{fer}. 
A general  problem in addressing the computation of SUSY-breaking  
soft masses is  the model-dependence of the results. In 
\cite{il,kl,bim} a slightly more model-independent approach was suggested
\footnote{An analogous analysis for compact Type IIB orientifolds was
given in \cite{imr}.}.
In most heterotic compactifications there are chiral fields corresponding 
to the complex dilaton $S$ and K\"ahler moduli $T^i$, whose (Planck mass 
suppressed) couplings to the charged fields are rather model-independent.  
The idea is to assume that the corresponding auxiliary fields $F^S$, $F^i$ 
get a vev, breaking SUSY. Even without knowing the microscopic source of 
this SUSY-breaking, the model-independence of the couplings of these 
fields allows to obtain specific predictions for soft terms, under
the assumption of a vanishing cosmological constant.
On the other hand a shortcoming of this approach is
precisely that the specific mechanism  of SUSY-breaking remains 
unexplained.
In addition the prediction for the scalar soft masses does depend on 
the tuning of the cosmological constant.

\medskip

In the last five years there have appeared several new ingredients in 
string theory which warrant a revision of these old problems. Specific 
Type II and Type I compactifications have been built with realistic 
particle spectrum. In these models the SM fields live in the worldvolume 
of D-branes embedded in a 10d closed string background 
\cite{aiq,aiqu,jei,more,inter,interrev}. 
Concerning moduli 
stabilization, it has been shown that e.g. compactifications of type IIB 
(on Calabi-Yau orientifolds) with non-trivial backgrounds of NSNS and RR 
3-form field strength fluxes \cite{fluxes} 
generically fix the vevs of the dilaton, and 
all complex structure moduli \cite{dgs,gkp,fp,kst,tt,ferrara,kklt,kors}.
 Thus the status concerning dilaton/moduli 
stabilization (in these and other compactifications with fluxes) has 
substantially improved. Unfortunately, no similar progress has been 
achieved concerning our understanding of the cosmological constant, which 
is still assumed to be tuned (in a discrete manner, e.g. by choices of 
flux quanta) to a small value.
 
In the present article we address the question of the structure of  
SUSY-breaking soft terms in Type IIB models with the charged fields  
living on the volume of D3-branes. The idea is to explore the structure 
of these soft terms, in an as model-independent way as possible.
We consider D3-branes located at a smooth (or orbifold) point in compact   
space, in a quite general closed string background. The 
non-trivial background (for instance, the presence of NSNS and RR 
3-form fluxes) gives rise in general to SUSY-breaking soft terms on the 
D3-brane worldvolume theory, which can be computed by coupling the closed 
string supergravity background to the D3-brane worldvolume theory using 
the Dirac-Born-Infeld (DBI) and Chern-Simons (CS) D3-brane action
\footnote{
This approach was already applied to a particular orbifold example in 
\cite{grana}.}.
It is 
important to emphasize that, in order to obtain the soft terms, only 
{\it local information } of the background around the position of the branes is 
required. Thus we will perform an expansion of the closed string 
background around the positions of the branes, and obtain the 
corresponding expansion of soft terms. Namely, to lowest order 
one generically (unless the D3-branes happen to sit at a singularity) 
gets the $\NN = 4$ super Yang-Mills action. Subsequent terms in the Taylor 
expansion of the supergravity background induce new operators on the 
worldvolume field theory, which break softly the $\NN = 4$ supersymmetry. 
The same technique may be applied to the case of the D3-branes sitting at 
an orbifold singularity, where the gauge sector may contain chiral 
fermions (and be phenomenologically interesting), and the background 
breaks softly the $\NN=1$ supersymmetry of the worldvolume theory. 

The structure of these soft terms depends on the structure of the 
underlying closed string background. For instance, they  depend on the
tensor components of the background 3-form flux $G_{ijk}$. We find that in 
some cases one obtains soft terms of the kind already found in the past in 
heterotic   
compactifications. Thus, as remarked in \cite{gkp}, an imaginary 
self-dual (ISD) 3-form flux 
leads to no soft terms, corresponding to the no-scale 
\cite{ns} (or modulus 
dominance) models. On the other hand  we find that an imaginary 
anti self-dual (IASD) $(3,0)$ background flux gives rise to soft terms 
corresponding to the 'dilaton domination' of old \cite{il,kl,bim}. More 
general tensor structures for $G_{ijk}$ give rise to more general 
possibilities.
The method also applies in backgrounds with other sources for 
SUSY-breaking, for instance it allows to compute the soft terms induced on 
the D3-branes from other distant sources like e.g. anti-D3-branes and 
orientifold O3-planes. In these examples, which involve no 3-form
flux, only scalar masses (no gaugino masses) are 
induced.

The D3-brane system in a local closed string background may be embedded 
into a complete compactification, so the structure of soft terms that we 
obtain may be present in a variety of Type IIB compactifications. In 
particular, the system may be embedded in a Calabi-Yau orientifold or 
F-theory 
compactification, such as those considered in \cite{gkp}. This 
provides an 
alternative derivation of the soft terms, by using the $\NN=1$ 
supergravity formalism, which allows to identify some of the particular 
background data (like the 3-form fluxes) with auxiliary components of 
certain supermultiplets (like the dilaton, and K\"ahler and complex 
structure moduli). This procedure makes contact with the one advocated in 
\cite{il,kl,bim} for the heterotic compactifications.
However, in the present Type IIB analysis there is an
explicit origin for SUSY-breaking, which is the underlying
supergravity background which includes 3-form fluxes.

We also observe that the $\NN=1$ supergravity effective action, with a
superpotential induced by  $G_{ijk}$ fluxes, is invariant under the
$SL(2,{\bf Z})$ S-duality for the complex dilaton \cite{filq} 
directly inherited from
Type IIB S-duality. The corresponding soft terms are thus also S-duality
invariant.

As we remark in the main text, 
the soft terms corresponding to dilaton domination have remarkable 
finiteness properties \cite{im,finite2,strings95}.
 Indeed, an $\NN=4$ theory with the addition of this 
class of soft terms remains perturbatively finite
\cite{finite1,finite2,finite3,finite4}. The reason for this 
fact was never very clear when this observation was made. In the present 
context we show that the finiteness  properties are  a consequence of 
holography and the fact that the IASD $(3,0)$ $G_{ijk}$ background has a 
constant dilaton.

As we have emphasized, the method presented to compute soft terms may be 
applied to explicit compact models. A large class of type IIB Calabi-Yau 
or F-theory compactifications was introduced in \cite{dgs,gkp}, with 
O3-planes, 
D3-branes and ISD fluxes. In that class of models the fluxes are ISD, and 
soft terms are absent on the worldvolume of $D3$-branes due to the 
no-scale structure. On the other hand we point out that locating 
$\ov{\rm D3}$'s on such ISD backgrounds does give rise to interesting 
SUSY-breaking soft terms on the worldvolume of the $\ov{\rm D3}$'s.
As an application of these ideas we show an specific example based on a 
$\IZ_3$ Type IIB orientifold model with $\ov{\rm D3}$-, D7-branes and 
antibranes. The set of branes at the $\IZ_3$ singularity gives rise to a 
realistic left-right symmetric $SU(3)\times SU(2)_L\times SU(2)_R\times 
U(1)_{B-L}$ three generation model. The underlying ISD flux gives rise
to soft terms of the dilaton dominated type on the worldvolume of the 
$\ov{\rm D3}$-branes and at the same time helps in cancelling 
RR-tadpoles.

Another interesting application concerns the 
structure of mass scales in the theory. In particular, one would like to 
identify the soft terms with those required by the MSSM for its 
viability. We find that soft masses are of order $M_s^2/M_p$, with $M_s$ 
the string scale. In order to obtain soft terms of order the electroweak 
scale one thus needs to have $M_s=10^{10}$ GeV. As described in the text, 
this is compatible with the observed value of the Planck scale if the 
compactification radius is three  orders of magnitude smaller than the 
string scale. Alternatively, if some large warp factor is present, the 
smallness of the weak scale could be due to the warping and still have 
$M_s$ of order the Planck scale.

The structure of the present paper is as follows. In section \ref{dbics} 
we describe the computation of the general soft terms by expanding the 
D3-brane worldvolume action in a general supergravity background. In 
section \ref{pureflux} we compute the soft terms on a D3-brane in a 
general background of 3-form fluxes. Several important classes of flux 
backgrounds, and the corresponding soft terms are discussed in section 
\ref{examples}. In section \ref{antibranes} we discuss soft terms arising 
from supersymmetry breaking by distant anti-D3-branes. In section 
\ref{dseven} we comment on the different pattern of soft terms expected 
on the worldvolume of  D7-branes. In section 
\ref{orbifolds} we show how to describe soft terms for D3-branes at 
orbifold singularities, in particular in examples leading to a chiral 
worldvolume spectrum, and give some examples. In section \ref{sugra} we 
describe the effect of flux-induced  SUSY-breaking from the viewpoint of 
the effective 4d $\NN=1$ supergravity action, and identify the auxiliary 
fields acquiring vevs with different components of the background flux. 
We also discuss some interesting properties of soft terms, like their 
S-duality invariance, and the relation between finiteness and dilaton 
domination via holography. In section \ref{realistic} we apply our 
results to the computation of soft terms in a semirealistic 3-family 
left-right symmetric extension of the Standard Model. Some 
phenomenological implications of our results are described in section 
\ref{pheno}, and we leave some conclusions and comments for section 
\ref{conclu}.

\section{SUSY-breaking soft terms  and fluxes in non-compact spaces}
\label{dbics}

In this section we describe the computation of soft terms 
\footnote{See appendix A for a short review of $\NN=1$ SUSY-breaking
soft terms and their ultraviolet properties.} from the 
perspective of the worldvolume field theory on a D3-brane in a quite 
general IIB supergravity background.
To do that we Taylor expand the supergravity background around 
the location of the branes. The transverse coordinates correspond to the
worldvolume scalars of the branes. Thus the first orders in this
expansion will yield the lowest dimensional  terms we are interested on.

\subsection{General strategy}

In a generic situation, the dynamical fields on the worldvolume of a 
stack of $N$ D3-branes are $U(N)$ gauge bosons, four adjoint Majorana 
fermions $\psi^a$, and six real adjoint scalars $\phi^m$.

In principle we should take the D3-brane action, given by a 
Dirac-Born-Infeld (DBI) term and a Chern-Simons (CS) term. Our conventions 
are described in appendix \ref{conventions}. Following \cite{myers} the 
Dirac-Born-Infeld piece
\footnote{There is the 
familiar question of the non-abelian extension of the DBI action. For the 
soft terms to be discussed in the present paper, only the leading 
$\alpha'$ order terms are relevant, for which no ambiguity arises.}, 
is given by 
\beqa
S_{D3}\, = \, -\mu_3 \int d^{4}x \,\Tr\left(e^{-\phi}\sqrt{-\det\left(
P\left[E_{\mu\nu}+E_{\mu m}(Q^{-1}-\delta)^{mn}\,E_{n\nu}\right]+
\sigma\,F_{\mu\nu}\right)\, \det(Q)} \right)\  \quad \quad
\label{dbi}
\eeqa
where $P[M]$ denotes the pullback of the 10d background $M$ onto the 
D3-brane worldvolume, and
\beqa
 E_{MN} & = & G_{MN}-B_{MN} \nonumber\\ 
 Q^m{}_n & = & \delta^m{}_n+i\sigma\,[\phi^m,\phi^p]\,E_{pn}\\
 \sigma & = & 2\pi \alpha' \nonumber
\eeqa
The Chern-Simons piece is given by
\beqa
S_{CS}=\mu_3\int \Tr\left(P\left[e^{i\sigma\,\hi_\phi \hi_\phi}\left(\sum_n 
C^{(n)}+\frac{1}{2}B_2\wedge C_2\right)e^{-B}\right]
e^{\sigma\,F}\right)\ .
\label{cs}
\eeqa
where $i_\phi C_{(p)}$ denotes contraction of a leg of the $p$-form, 
transverse to the D3-brane, with the associated worldvolume scalar 
\cite{myers}. The above action is the bosonic sector of the full 
supersymmetric action. The relevant piece of the fermionic sector will be 
discussed later on.

The relative sign between the DBI and CS terms corresponds to the relative 
sign between the tension and charge of the 3-brane, and distinguishes 
D3-branes and anti-D3-branes (denoted $\ov{\rm D3}$-branes in what 
follows).

The above action reduces at low energies (i.e. at leading order in 
$\alpha'$) to the action of $\NN=4$ super Yang-Mills
(plus a constant piece corresponding to the tension). In particular the 
DBI piece gives rise to  the gauge field and scalar kinetic terms, as well as 
the familiar commutator squared potential for the scalars, as we see below. 
The axion coupling from the CS terms provides the $\theta$ term of 
Super-Yang-Mills (SYM). 

Notice that the familiar $SO(6)$ R-symmetry of $\NN=4$ SYM (under which 
the worldvolume scalars and fermions transform in the representations 
{\bf 6}, {\bf 4}, respectively) is also present in the full D3-brane 
action. In fact, it corresponds to the geometric rotational symmetry in 
the space transverse to the D3-brane. As such, it is a local symmetry of 
the configuration, which constrains the possible coupling of D3-brane 
worldvolume fields with the 10d supergravity fields.

\medskip

It is important to recall the interpretation of worldvolume scalars as 
coordinates in transverse space, via
\beqa
x^m=2\pi\alpha'\, \phi^m \ \ \ .
\eeqa
This implies that for the 10d supergravity fields in the above action,  
the dependence on the coordinates $x^m$, $m=4,\ldots, 9$ is translated, 
from the viewpoint of the D3-brane worldvolume, into a dependence on the 
worldvolume scalars. This leads to a nice interpretation, as follows. 
One can expand the supergravity background around the trivial 
configuration (flat metric, constant dilaton, vanishing NSNS and RR 
background) as a power series in $x^m$. This is clearly an expansion 
around the location of the D3-branes, so that the lowest terms provide a 
good description of the local background felt by the D3-branes. There is 
also a corresponding expansion on the D3-brane worldvolume. Regarding the 
10d background fields as expansions in powers of $\phi^m$, one can expand 
the action (\ref{dbi}), (\ref{cs}) around the $\NN=4$ SYM theory, finding 
an infinite set of terms, of higher and higher dimensions, deforming 
the original $\NN=4$ theory. The $\alpha'$ expansion in the D3-brane 
worldvolume, therefore corresponds to an expansion around the local 
geometry near the D3-branes. The truncation of the D3-brane effective 
field theory to operators of some dimension corresponds to using a local 
description of the supergravity background on which the D3-branes are 
embedded, to some order in the power expansion.

From any of these two physical points of view, we are interested in the 
first few such terms. From the D3-brane worldvolume viewpoint, this 
corresponds to keeping the most relevant terms; we will restrict to terms 
of dimension at most 3 in the computation below, although more terms can 
be obtained using this technique. From the viewpoint of the 10d geometry,
this corresponds to taking a local description of the configuration. 
This is interesting since it is in general easier to analyze the behaviour 
of local geometries, and later on embed them in global compactifications. 
In this paper we center on quite general configurations of supergravity 
fields for D3-branes at a smooth point in non-compact Calabi-Yau 
threefolds (which locally correspond to D3-branes in flat space, or 
rather, flat except for the backreaction of the fluxes on the metric). 
We also discuss the case of D3-branes sitting at orbifold singularities,
which lead to chiral theories and could have phenomenological relevance.
These classes  provide  local models  for a large set of IIB string 
compactifications. Finally, another important application of local 
configurations of D3-branes is the AdS/CFT 
correspondence, since only the local geometry around the D3-branes 
survives in the near-horizon limit. In this language, the computation of 
soft terms for a given background corresponds to the computation of the 
mapping between low-dimension operators in the gauge theory and the 
supergravity fields in its gravity dual.

An important comment is in order. In our computation we expand around a 
configuration of D3-branes. It is possible that the induced soft terms 
lead to some instability of this original configuration. One prototypical 
example is the appearance of linear terms (see below), or of instabilities 
triggered by scalar trilinear terms, which induce the appearance of 
non-commuting scalar vevs, and hence the polarization of the D3-branes 
into spherical 5-branes \cite{myers,ps}. 
In such unstable situations, our computation is useful in that it shows
the instability, but one would need to carry out the analogous 
expansion around the stable vacuum to correctly describe the physical 
system. However in many situations, as we will see later, there are 
scalar mass terms which either remove the polarized D3-brane minimum, or 
create a potential barrier to reach it, thus rendering the original 
configuration at least metastable. Thus our analysis applies without 
subtleties to these situations. 

\subsection{The soft terms}

In this paper we will center on quite general backgrounds which preserve 
4d Poincare invariance, and include objects like D3-branes, antibranes, 
and 3-form fluxes in the transverse space. This implies some restriction, 
for instance we will not allow for e.g. a constant term in the expansion 
of the 5-form (i.e. we do not describe 5-form fluxes, but include 5-form 
backgrounds generated by sources). We will however find that this class 
of background has a lot of interesting physics. The supergravity 
backgrounds we consider follow the general ansatz
\beqa
ds^2 & = & Z_1(x^m)^{-1/2} \eta_{\mu\nu}\, dx^\mu dx^\nu + Z_2(x^m)^{1/2}
\, ds^2{}_{CY} \nonumber \\
\tau & = & \tau(x^m) \nonumber \\
G_3 & = & \frac{1}{3!}G_{lmn} (x^m)\, dx^l dx^m dx^n \nonumber \\
\chi_4 & = & \chi(x^m) \, dx^0 dx^1 dx^2 dx^3 \\
F_5 & = & d\chi_4 + *_{10}\, d\chi_4  \nonumber
\label{ansatz}
\eeqa
where $G_3=F_3-\tau H_3$ (with $F_3(H_3)$  being the RR(NSNS) flux),  
and $ds^2{}_{CY}$ denotes the metric in 
transverse space, in the absence of flux backreaction, i.e. the 
Ricci-flat metric of the underlying Calabi-Yau. Since we are interested 
in a local description, it is enough for our purposes to take it to be 
flat 6d metric. 

\medskip

In general, the above fields will be subject to the supergravity 
equations of motion, with sources. For the moment we do not impose the 
equations of motion, since they will depend on the particular sources in 
the configuration.

The fields in general will have a power expansion
\beqa
Z_1^{-1/2} & = & 1 + \frac{1}{2} K_{mn}\, x^m x^n + \ldots \nonumber \\
Z_2^{1/2} & = & 1 + \ldots \nonumber\\
\tau & = &  \tau_0 + \frac{1}{2}\tau_{mn}\, x^m x^n \nonumber \\
\chi_4 & = & ({\rm const.} +\frac{1}{2}\chi_{mn}\, x^mx^n+\ldots) \,
dx^0 dx^1 dx^2 dx^3 \\
G_{lmn} (x^m)  & = & G_{lmn} + \ldots \nonumber 
\label{powerexp1}
\eeqa
where the coefficients $K$, $F$, $G$, $\tau_0$ in the right 
hand side are constant, independent of $x^m$. The piece of the 5-form 
background relevant for our purposes below can be given also as
\beqa
F_5 & = &  
 \frac{1}{2}(\chi_{mn}+\chi_{nm}) x^m dx^n dx^0 dx^1 dx^2 dx^3  
+\ldots
\label{powerexp2}
\eeqa
The order of expansion will be enough for our purposes, but it should be 
clear that the techniques below can be similarly applied to obtain 
higher-dimensional terms. The above expansion and the corresponding 
expansion in 
the D3-brane action are a generalization of that studied in 
\cite{grana} for D3-branes at an orbifold singularity. We will discuss 
the orbifold situation in section \ref{orbifolds}, and see that it is a 
simple projection of our more general result. 

A comment is in order. In general, the above expansions for e.g. the 
metric or dilaton may  include terms linear in the coordinates. Following 
computations similar to the ones below, they in general lead to 
worldvolume field theory terms linear in the scalar fields $\phi^m$. 
This implies (unless a fine-tuning of terms of different origin occurs, 
so that the linear terms cancel) that the D3-branes are not sitting at an 
extremum of the worldvolume scalar potential, and therefore that the 
latter acquire vevs and run down the potential. Geometrically, the 
D3-branes are not sitting at a stable location, and tend to move and fall 
into e.g. gravitational potential wells. In appendix \ref{linear} we 
discuss an explicit example of this situation, where a D3-brane in the 
supergravity background of a stack of anti D3-branes falls towards them. 
From the viewpoint of the worldvolume theory, the background induces 
soft terms including linear terms in the scalars, which forces them to 
acquire vevs. Since this kind of situation is unstable, in the above 
expression we assumed just terms that lead to scalar potentials starting 
at quadratic order.

\medskip

In what follows we describe the lowest order terms arising from the above 
expansion in the action (\ref{dbi}), (\ref{cs}). 
Starting with the DBI action, we obtain the following expansion for the 
different terms (notice that $E_{\mu n}=0$ in our ansatz)
\beqa
\pim \,\tau & = & \pim \,\tau_0 + \frac{\sigma^2}{2} \pim\, \tau_{mn}\, 
\phi^m \phi^n  \nonumber \\
\det(Q)^{1/2} & = & 1 - \frac{i\sigma}{2} \, B_{mn} \, [\phi^n,\phi^m] \, 
-\, \frac{\sigma^2}{4} \, [\phi^m,\phi^n] \, [\phi_m,\phi_n] \nonumber \\
\, [ -\det ( P [E_{\mu\nu}] ) ]^{1/2} & = & 1 - 
\frac{\sigma^2}{2}\, \partial_\mu \phi^m\, 
\partial^\mu \phi^m \,+\, \sigma^2\,  K_{mn}\, \phi^m\phi^n
\label{determ}
\eeqa
Grouping things together, we obtain 
\beqa
{\cal L}_{\rm DBI} & = & \frac{\mu_3\sigma^2}{g_s} \Tr \left[ \right. 
\frac{1}{2}\, \partial_\mu \,\phi^m \partial^\mu \phi^m -\, \frac{1}{4} \, 
[\phi^m,\phi^n] \, [\phi_n,\phi_m]- \nonumber\\
& & \,-\, (K_{mn}+ \frac{g_s}{2} \pim\, \tau_{mn})\, \phi^m\phi^n 
-\frac {1}{12} \, g_s \, (G_3-{\ov G}_3)_{lmn}\, \phi^l\, [\phi^n,\phi^m]
\left. \, \right]
\label{softdbi}
\eeqa
where we have used that $B_{mn}=\frac{1}{3}H_{lmn}\, x^l= 
\frac{1}{3}\, \sigma\, H_{lmn}\phi^l
=\frac{1}{6}\, \sigma \, i \, g_s(G_3-{\ov G}_3)_{lmn} \,\phi^l$.
Eq. (\ref{softdbi}) corresponds to the familiar bosonic lagrangian of 
$\NN=4$ SYM, in the first line, and some soft terms in the second.

It is clear that in general, the expansion of the background, and its 
corresponding expansion of the D3-brane action, can contain terms leading 
to additional cubic and quartic couplings in the scalars. About the 
former, they will be absent in our examples below, essentially due to 
symmetries of the background ansatz. About the latter, we ignore them 
since we center on at most dimension 3 soft terms. In any case, these 
terms can be straightforwardly introduced in the general expressions, as 
the interested reader may check. 

Carrying out a similar expansion in the CS piece, we have several terms
beyond the familiar $\NN=4$ D3-brane CS action, involving
\beqa
(i_{\phi} i_{\phi} C_6)_{0123} & = & \frac {\sigma}{2} 
(C_6)_{0123mn} [\phi^n,\phi^m] \nonumber\\
(C_4)_{0123} & = & \frac{\sigma^2}{2}\chi_{mn}\, \phi^m \,\phi^n
\eeqa

Using that in our conventions $C_6$ is defined \cite{ps} through (\ref{relationc6})
\beqa
-*_{10}(F_3-CH_3)=dC_6-H_3\wedge \left(C_4+\frac{1}{2}B_2\wedge C_2\right)
\eeqa   
and the fact that at leading order in the scalars $C_4$ is trivial, we 
have $(dC_6)_{0123mnp}= -\frac{1}{2}[*_6(G_3+{\ov G_3})]_{mnp}$, and 
$(C_6)_{0123mn}=-\frac{\sigma}{6} [*_6(G_3+{\ov G_3})]_{mnp} 
\phi^p$.

The CS piece of the action hence reads
\beqa
{\cal L}_{CS}\, =\, \mu_3 \sigma^2\, \Tr \left[\, \frac 12 (\preal \tau)\, 
F_{\mu\nu} {\widetilde F}^{\mu\nu}\, +\,\frac{1}{2}\, \chi_{mn}\, 
\phi^m\phi^n\, -\, \frac {i}{12}\, *_6(G_3+{\ov G_3})_{lmn} \, \phi^l 
[\phi^n,\phi^m] \, \right]
\quad\quad
\eeqa

\medskip

Finally we should also carry out a similar expansion in the fermionic 
piece of the action to get the soft fermion masses (the only fermionic 
soft terms of dimension up to three). This can be obtained by 
dimensional reduction of the fermionic completion of the 10d 
supersymmetric DBI-CS action \cite{dbifer}. This expansion in powers of 
the fermionic fields has been worked out by different methods (see for 
example \cite{grafer},\cite{norcor}). In particular, here we are interested 
in the piece giving rise to fermionic masses and kinetic terms
\beqa
{\cal L}_{\rm ferm.}\,  =\, 
\frac{\sigma^2\mu_3}{g_s}\, \left(\, -\frac{1}{2}\, \bar{\Theta}\, 
\Gamma^{\mu} \,D_{\mu}\,\Theta + \frac{1}{48}g_s\, \bar{\Theta}\, 
\Gamma^{pqr}\, \Theta 
Re[(a+ib)^2(*_6G-iG)_{pqr}]
\right)
\eeqa
Here the kinetic terms indices are 4d, while the indices $p,q,r$ are 10d 
indices, but running only over the directions transverse to the D3-branes, 
since only on these there is non-trivial flux background. 
Also $\Theta$ is the ten-dimensional Majorana-Weyl spinor formed by the 
massless Ramond open string states, and $a$ and $b$ fix the embedding of 
the D3-brane 
supersymmetry in the 10d $\NN=2$ IIB supersymmetry by means of 
$\theta_1 = a\sigma\Theta$, $\theta_2 = b\sigma\Theta$, where $\theta_1$, 
$\theta_2$, are the two Majorana-Weyl spacetime spinors of $\NN=2$ 
supergravity and $a^2+b^2=1$. We work in the choice $(a,b)=(1,0)$.

Performing the reduction to four dimensions according to
\beqa
\Gamma^{\mu}  = \gamma^{\mu}\otimes 1 \quad \quad ; \quad\quad
\Gamma^m  = \gamma_{(5)} \otimes \gamma^m
\eeqa
we get the soft term fermionic lagrangian:
\begin{eqnarray}
{\cal L}_{\rm ferm.}\, =\, 
\frac{\mu_3\sigma^2}{g_s}\biggr(-\frac{1}{2}\sum_a\bar{\psi^a}\,
\gamma^{\mu}\,D_{\mu}\psi^a + \frac{g_s} {96} 
(*_6G_3-iG_3)_{mnp}\, \sum_a\psi^a\, (\gamma^{mnp})_{ab}\,\psi^b + 
{\rm h.c.}\biggr) \quad\quad
\end{eqnarray}
Here the Dirac matrices act on the $SO(6)$ weight space corresponding to 
the four different Majorana fermions on the D3-brane, labeled by 
$a,b,\ldots$. Notice that there 
is no Dirac matrix acting on the 4d spacetime spinor indices, so this is 
indeed a mass term.
The Yukawa couplings of the $\NN=4$ theory can be obtained by considering 
higher-dimensional terms in the expansion.

Notice that the result can be easily guessed, since it is the only 
possible $SO(6)$ invariant mass term. Namely, it contracts the imaginary 
anti self-dual part of the flux (which transforms in the representation 
${\bf 10}$ of $SO(6)$) with the symmetric product of two fermions in the 
${\bf \ov{4}}$ (which transforms in the ${\bf \ov{10}}$ of $SO(6)$), see 
below.

\medskip

In total, we have the soft term lagrangian, in the string frame,
\beqa
{\cal L}_{\rm soft} & = & \frac{\mu_3\sigma^2}{g_s}\Tr \left[ 
\, -\, (\,K_{mn}-\frac{g_s}{2}\chi_{mn}+\frac{g_s}{2} \pim \tau_{mn}\,) 
\,
\phi^m\phi^n + \right. \\ 
&+ & \left. 
\frac {ig_s}{6}\, (*_6 G_3-iG_3)_{lmn}\, \phi^l\phi^m\phi^n\, +\, 
{\rm h.c.}\, +\, \frac{g_s}{96}\, (*_6 G_3-iG_3)_{lmn} \, \psi 
\, \gamma^{lmn} \, \psi\, +\, {\rm h.c.} \right] \nonumber
\eeqa
This may be rewritten in the Einstein frame by means of 
$G_E=g_s{}^{-1/2}G_{str}$ and $\gamma^M_E=g_s{}^{1/4} \gamma^M_{str}$. 
Rescaling all the fields in order to get rid of the 
$\mu_3 \sigma^2 = \frac{1}{2\pi}$ global factor and the fermions in order
to get the usual factor $i$ in the kinetic term, it takes the form
\beqa
{\cal L}_{\rm soft} & = & \Tr \left[ \, -
\, (\,K_{mn}-\frac{1}{2}\chi_{mn}+\frac{g_s}{2} \pim \tau_{mn}\,) 
\phi^m\phi^n \, + \right. \\ 
&   & \left. +\frac {ig_s\sqrt{2\pi}}{6}\, (*_6 G_3-iG_3)_{lmn}\, 
\phi^l\phi^m\phi^n\, +\, {\rm h.c.}\, +\, \frac{ig_s^{1/2}}{96}\,
 (*_6 G_3-iG_3)_{lmn}\, \psi \, \gamma^{lmn}\, \psi\, +\, {\rm h.c.} \,
\right] \nonumber 
\label{finalsoft}
\eeqa
This is the main result of this section, the soft term lagrangian for a 
stack of D3-branes in a (quite) general supergravity background, of the 
form (\ref{ansatz}).

\medskip

For D3-branes in Calabi-Yau manifolds, there is a preferred complex 
structure, associated to the $\NN=1$ supersymmetry unbroken by the 
Calabi-Yau and the D3-brane. Namely, the complex structure in which the 
$SU(3)$ invariant spinor satisfies $\gamma^{\ov {\i}}\xi=0$. It is useful 
to combine the D3-brane worldvolume fields in a vector multiplet and three 
chiral multiplets under this $\NN=1$ supersymmetry, and express the soft 
terms in complex notation. This also simplifies the comparison with the 
familiar supersymmetry breaking terms in $\NN=1$ supersymmetric field 
theories.

For concreteness, take $z^1=\frac{1}{\sqrt{2}}(x^4+ix^5)$, 
$z^2=\frac{1}{\sqrt{2}}(x^6+ix^7)$, $z^3=\frac{1}{\sqrt{2}}(x^8+ix^9)$, and
introduce the complex scalars $\Phi^1=\frac{1}{\sqrt{2}}(\phi^4+i\phi^5)$,
$\Phi^2=\frac{1}{\sqrt{2}}(\phi^6+i\phi^7)$, $\Phi^3=\frac{1}{\sqrt{2}}
(\phi^8+i\phi^9)$. We also use $\Phi^{\ov {\i}}$ to denote $\Phi_i^*$.
Denoting the four-plet of fermions by their $SO(6)$ weights, the fermion 
$\frac 12(+++)$ belongs to the vector multiplet, and is referred to as the 
gaugino, denoted $\lambda$. The fermions combining with the 
above complex scalars to give chiral multiples are $\Psi^1$, $\Psi^2$, 
$\Psi^3$, corresponding to the weights $\frac 12(+--)$, $\frac 12(-+-)$ 
and $\frac 12 (--+)$, respectively.

In a preferred complex structure, only an $SU(3)$ (times $U(1)$) subgroup 
of the $SO(6)$ symmetry is manifest. The antisymmetric flux $G_{mnp}$
transforms as a 20-dimensional reducible $SO(6)$ representation, 
decomposing as $20={\ov{10}}+10$. The irreducible representations 
${\ov{10}}$, $10$ correspond to the imaginary self-dual (ISD) $G_{(3)}^+$ 
and imaginary anti self-dual (IASD) $G_{(3)}^-$ parts, respectively, defined as
\beq
G_{(3)}^{\pm}\ =\ {1\over 2} (G_{(3)}\mp i*_6 G_{(3)})
\ \ ;\ \  *_6G_{(3)}^{\pm}\ =\ \pm iG_{(3)}^{\pm}
\label{ISD}
\eeq
It is useful to classify the 
components of the ISD and IASD parts of $G_3$  according to their 
behaviour under $SU(3)$, $10=6+3+1$. For that purpose, we introduce the tensors
\cite{gp}
\beqa
S_{ij} &=& 
\frac{1}{2}(\epsilon_{ikl}G_{j\bar{k}\bar{l}}+
\epsilon_{jkl}G_{i\bar{k}\bar{l}}) \nonumber\\
A_{\bar{i}\bar{j}} &=&
\frac{1}{2}(\epsilon_{\bar{i}\bar{k}\bar{l}}G_{kl\bar{j}}-
\epsilon_{\bar{j}\bar{k}\bar{l}}G_{kl\bar{i}}) \nonumber
\eeqa
defined in terms of the complex components of $G_3$, and which transform 
in the representation ${\bf 6}$ and ${\bf 3}$ under $SU(3)$.
One similarly defines $S_{\bar{i}\bar{\j}}$ and $A_{ij}$.

The ISD condition allows for a $(0,3)$ component, a $(2,1)$ primitive part
\footnote{ $G_3$ is primitive if  $G_3\wedge J=0$ with J the K\"ahler 
form.}
which corresponds to $S_{\bar{i}\bar{j}}$, and a $(1,2)$ non-primitive 
part corresponding to $A_{ij}$. The IASD condition allows for a $(3,0)$ 
part, a $(1,2)$ primitive part ($S_{ij}$) and a $(2,1)$ non-primitive one 
($A_{\bar{i}\bar{j}}$). The corresponding $SU(3)$ representations are 
given in  table 1.

\begin{table}[htb]
%\footnotesize
\renewcommand{\arraystretch}{0.70}
\begin{center}
\begin{tabular}{|c|c|c||c|c|c|}
\hline      & ISD   &     &   & IASD &   \\
\hline   $SU(3)$ rep. &  Form  &  Tensor &  $SU(3)$ rep. & Form & Tensor
\\
\hline \hline
 ${\overline 1}$  & $(0,3)$  &  $G_{{\bar 1}{\bar 2}{\bar 3}}$ &
 $ 1$  & $(3,0)$  &  $G_{{ 1}{ 2}{ 3}}$ \\
\hline
 ${\overline 6}$  & $(2,1)_P$  &  $ S_{{\bar i}{\bar j}}$ &
 ${ 6}$  & $(1,2)_P$  &  $ S_{{ i}{ j}}$ \\
\hline
${\overline 3}$  & $(1,2)_{NP}$  &  $ A_{{ i}{ j}}$ &
 ${ 3}$  & $(2,1)_{NP}$  &  $ A_{{\bar  i}{\bar j}}$ \\
\hline \end{tabular}
\caption{$SU(3)$ decomposition of antisymmetric $G_{(3)}$ fluxes.}
\end{center}
%\label{cogno}
\end{table}

Using the complex notation, and these definitions, we have the (Einstein 
frame) soft term lagrangian 
\beqa
{\cal L}  =  \Tr \biggr[ &
-\, (\,2K_{i\bar{\j}}-\chi_{i\bar{\j}}+g_s (\pim \tau)_{i\bar{\j}}\, )\,
\Phi^i\Phi^{\bar{\j}}\, -\, \frac{1}{2}\,(\, 2K_{ij}-\chi_{ij}+g_s(\pim 
\tau)_{ij}\, )\, \Phi^i\Phi^j\, +\, {\rm h.c.}  \nonumber \\
& +g_s\sqrt{2\pi}\, \left[\, \frac{1}{3}G_{123}\, \epsilon_{ijk}\, 
\Phi^i\Phi^j\Phi^k + \frac{1}{2}\epsilon_{\bar{i}\bar{j}\bar{l}}\, 
(S_{lk}-(A_{\bar{l}\bar{k}})^*)\, \Phi^{\bar{i}}\Phi^{\bar{j}}\Phi^k\, +\, 
{\rm h.c.} \, \right]\, + \nonumber \\
&+\, \frac{g_s^{1/2}} {2\sqrt{2}} \, \left[ \, G_{123}\, \lambda\lambda 
\, +\, \frac{1}{2}\epsilon_{ijk}A_{\bar{j}\bar{k}}\, \Psi^i\lambda \, +\,
\frac{1}{2}S_{ij}\, \Psi^i\Psi^j \, +\, {\rm h.c.}\, \right]\,\, \biggr]
\label{softgencomp}
\eeqa
where we have defined
\begin{eqnarray}
(\pim \tau)_{i\bar{\j}} = 
\frac{1}{2i}(\tau_{i\bar{\j}}-(\tau_{j\bar{i}})^*) \nonumber\\
(\pim\tau)_{ij} = \frac{1}{2i}(\tau_{ij}-(\tau_{\bar{\j}\bar{i}})^*)
\end{eqnarray}
It is standard to write the soft terms using the couplings of the 
underlying $\NN=1$ supersymmetric theory. In our case, one can write the 
$\NN=1$ superpotential of the $\NN=4$ theory as 
\beq
W_{\NN=4}\ =\ g{{\sqrt{2}}\over 3}\epsilon_{ijk} \Phi^i\Phi^j\Phi^k \ =\
\frac{1}{3!}h_{ijk} \Phi^i\Phi^j\Phi^k
\label{supn=4}
\eeq
with $g=\sqrt{2\pi g_s}$ the gauge coupling constant.
Thus, in the notation of Appendix A, we have soft SUSY
breaking  terms as follows:
\beqa
m_{ij}^2\ & =&\ 2K_{i\bar{\j}}-\chi_{i\bar{\j}}+g_s (\pim \tau)_{i\bar{\j}} 
\nonumber
\\
B_{ij}\ &=& \ 2K_{ij}-\chi_{ij}+g_s(\pim \tau)_{ij} \nonumber
\\
A^{ijk}\ &=& \ -h^{ijk} {{g_s^{1/2}}\over {\sqrt{2}}}\ G_{123}
 \nonumber
\\
C^{ijk}\ &=& +h^{ijl} {{g_s^{1/2}}\over {2\sqrt{2}}}(S_{lk}-(A_{{\bar l}{\bar 
k}})^*)
\nonumber
\\
M^a\ &=& \  {{g_s^{1/2}}\over {\sqrt{2}}}\ G_{123} \nonumber
\\
\mu _{ij} \ & =& \  -{{g_s^{1/2}}\over {2\sqrt{2}}} S_{ij} 
 \nonumber
\\
M_{g}^{ia}\ &=&\ {{g_s^{1/2}}\over {4\sqrt{2}}}
\epsilon_{ijk} A_{{\bar j}{\bar k}}
\label{softgen}
\eeqa
Several comments are in order. First note that all dimension 3 SUSY-breaking 
soft terms only depend on the three form flux, and not on the 5-form, 
warping or dilaton contributions. Note also that, even before applying
supergravity equations of motion, there are certain relationships 
between the different soft terms. In particular $\NN=1$ gaugino masses and 
$A^{ijk}$ trilinear terms are related by
\beq
A^{ijk}\ =\ - h^{ijk} M^a
\label{m=a}
\eeq
Note also that in certain situations the dimension 3 soft terms are 
consistent with one unbroken $\NN=1$ supersymmetry. In particular,
if there is only $(1,2)$ IASD flux (i.e. $S_{ij}\not= 0$), one
has the relationship 
\beq
C^{ijk}\  = \  -h^{ijl}\mu _{kl}
\label{sofn=1}
\eeq
which corresponds to the trilinear scalar term arising from an effective 
$\NN=1$ superpotential obtained by adding a mass piece 
 ${1\over 2}\mu_{ij}\Phi^i \Phi^j$ to the $\NN=4$ superpotential.
If in addition the scalars $\Phi^i$ and fermions $\Psi^i$ 
have equal masses the terms would preserve $\NN=1$.  We  
discuss these different situations, as well as additional relationships 
appearing after imposing locally the supergravity equations of motion 
below.

The result  eq.(\ref{softgencomp}) is more general than that in 
\cite{grana} in several respects. In particular  we do not 
impose orbifold symmetries. Hence our above expression contains some 
interesting new terms, like for  instance scalar mass terms of the form 
$\Phi^i\Phi^j$, commonly known as B-terms in the SUSY-breaking soft term 
literature, or non-holomorphic trilinear scalar terms.
Although such terms are projected in  e.g. abelian orbifold projections
of odd order, they are important whenever the orbifold allows for 
invariant $(1,2)$ forms (like for example $\IC^3/\IZ_4$).
 This situation is briefly discussed in section \ref{orbifolds}.

\medskip

Let us mention that the above result changes if one considers 
anti-D3-branes instead of D3-branes, via a flip in the relative sign 
between the DBI and CS pieces of the action. This implies an additional 
sign in the pieces related to $\chi_{mn}$ and $*_6 G_3$. This 
effectively exchanges the roles of the imaginary self-dual (ISD) and 
imaginary anti self-dual (IASD) parts 
of $G_3$, at the level of the soft term lagrangian. Hence the effect of 
ISD (resp. IASD fluxes) on D3-branes will be equivalent to the effect of 
IASD (resp. ISD) fluxes on $\ov{\rm D3}$-branes. In other words, a 
simultaneous parity in the 4d Poincare dimensions and the 6d transverse 
coordinates (which is a symmetry of type IIB theory) exchanges the roles 
of ISD and IASD fluxes and turns D3-branes into $\ov{\rm D3}$-branes, and 
vice versa.

\medskip

An additional comment is in order. The parameters in the supergravity 
background are 
in general constrained by the supergravity equations of motion. The 
explicit relations in different examples depend on the corresponding 
sources. We have chosen not to impose these constraints in the above 
expressions, in order to be completely general. In coming discussions and 
examples, however, we will ensure that the background obeys the 
supergravity equations of motion. Then the metric, dilaton and 5-form 
background are related to the corresponding sources, which we usually 
take to be just the fluxes themselves, as in next section. The above 
expressions, however, would be valid if other kinds of sources are 
present, see for instance section \ref{antibranes}.

\section{D3-branes in pure flux backgrounds and equations of motion}
\label{pureflux}

One very interesting general class of configurations are those where the 
(only) source of warping, dilaton and 5-form are 3-form fluxes.

The structure of the background is given by our above ansatz. As 
discussed, the supergravity equations of motion of course imply relations 
between the parameters in the ansatz for the metric, dilaton, and 5-form, 
and the 3-form fluxes that source them. 
In particular, we will find that equations of motion relate the 
form of mass squared terms to the 3-form fluxes.
The $D=10$ supergravity equations of motion 
(\ref{eomfirst}-\ref{eomcomplete}) lead to the following results.
From equation
\begin{equation}
\bigtriangledown^M \left(\, \frac{i}{2}\, e^{\phi} \,\partial_M \tau\, 
\right)\, -\, \frac{1}{2}\, e^{2\phi}\, \partial^M C\, \partial_M \tau\, =
\, \frac{g_s}{24}\, G_{MNP}G^{MNP}
\end{equation}
to lowest order we get the constraint
\begin{equation}
i\sum\tau_{l\bar{l}}\, =\, \frac{1}{2}\, \left(\, 
G_{123}G_{\bar{1}\bar{2}\bar{3}}\, +\, \frac{1}{4}\, 
S_{lk}S_{\bar{l}\bar{k}}\, +\, \frac{1}{4}\, A_{lk}A_{\bar{l}\bar{k}}\, 
\right) \end{equation}
From 
\begin{eqnarray}
R_{MN}&=& \frac{1}{4}\, e^{2\phi}\, \partial_M\tau \,\partial_N\tau^\ast\, 
+\, \frac{1}{4}\, e^{2\phi}\, \partial_M\tau^\ast\, \partial_N\tau \, +\,  
\frac{g_s^2}{96}F_{MQRST}F_N~^{QRST}
\nonumber\\
&&+\,\frac{g_s}{8}\, \left(\, {G_M}^{PQ}G_{NPQ}^\ast\, + \,
{G_N}^{PQ}G_{MPQ}^\ast\, -\, \frac{1}{6}\, g_{MN}\, G_{PQR}^\ast 
G^{PQR}\, \right)
\label{ricci}
\end{eqnarray}
we obtain
\begin{equation}
4\sum K_{l\bar{l}}=\frac{g_s}{2}\left(\vert G_{123}\vert^2 
+ \vert G_{\bar{1}\bar{2}\bar{3}}\vert^2 + 
\frac{1}{4}\sum_{ij}(\vert S_{ij}\vert^2 + 
\vert S_{\bar{i}\bar{j}}\vert^2 + \vert A_{ij}\vert^2 
+ \vert A_{\bar{i}\bar{j}}\vert^2)\right)
\label{tracemetric}
\end{equation}  
Finally from  
\begin{equation}
d {F}_5\, =\, \frac{ig_s}{2}\, G_3 \wedge \overline{G}_3
\end{equation}
we obtain the constraint 
\begin{equation}
-2\sum \chi _{l\bar{l}}\, =\, \frac{g_s}{2}\, \left(\, \vert 
G_{123}\vert^2 
\, -\, \vert G_{\bar{1}\bar{2}\bar{3}}\vert^2\, +\, 
\frac{1}{4}\, \sum_{ij}(\, \vert S_{ij}\vert^2 \, -\, 
\vert S_{\bar{i}\bar{j}}\vert^2\, -\, 
\vert A_{ij}\vert^2\, +\, \vert A_{\bar{i}\bar{j}}\vert^2\,)\, \right)
\end{equation}
Combining these results one finds the following constraint on the scalar 
mass matrix
\beqa
m_1^2 \, +\,  m_2^2\,  +\, m_3^2\, =\, 
\frac{g_s}{2}\, \left[ \, \vert G_{123} \vert^2\, +\,
\frac{1}{4}\, \sum_{ij}\, (\, \vert S_{ij} \vert^2\, +\, 
\vert A_{\bar{i}\bar{j}}\vert^2\, )\, -\, \nonumber \right. \\ 
\left. -\, \preal\, (\, G_{123}G_{\bar{1}\bar{2}\bar{3}}\, +\,
\frac{1}{4}\, S_{lk}S_{\bar{l}\bar{k}}\, +\, 
\frac{1}{4}\, A_{lk}A_{\bar{l}\bar{k}})\, \right] 
\label{genmass}
\eeqa
The first three terms in the right hand side may be rewritten 
in terms of the 
trace of the masses $M_{ij}$, $i,j=0,1,2,3$ of the four fermions
$\lambda$ and $\Psi_{1,2,3}$.  One then has
\beq
m_1^2 + m_2^2 + m_3^2 \, =\,  \Tr(M_{ij}(M^{ij})^*)\, - \,
\frac{g_s}{2}\, \preal\, (\, G_{123}G_{\bar{1}\bar{2}\bar{3}}+
\frac{1}{4}S_{lk}S_{\bar{l}\bar{k}}+\frac{1}{4}A_{lk}A_{\bar{l}\bar{k}}\, 
)
\label{genmass2}
\eeq
Notice that the supertrace of the masses of the scalars and spin 1/2 
fermions vanishes in certain cases. In particular it vanishes
if all the ISD fluxes vanish (i.e. $G_{\bar{1}\bar{2}\bar{3}}=
S_{\bar{l}\bar{k}}=A_{lk}=0$). It also vanish 
if each ISD(IASD) background is 
not present when its IASD(ISD) partner flux is present.

Notice that as observed in section IV.A of \cite{ps}, the masses are not 
fully determined by the fluxes, only the trace of the mass squared matrix 
is. This is because the masses can be changed (keeping the trace fixed) 
 by contributions from e.g. homogeneous vacuum solutions to the metric 
equation 
of motion. In appendix \ref{massesps} we describe one such situation. 
Since $B_{ij}$ terms do not contribute to the trace of the scalar masses,
the local equations of motion do not give any constraint on them
either.

In the case of the scalar mass terms $m_{ij}^2$ one can get 
a relationship with the fluxes 
 if a definite choice of boundary 
conditions for the metric, 5-form and dilaton backgrounds is made (which 
fixes the ambiguity of the vacuum solution). 
For example, assuming spherically 
symmetric boundary conditions, which are natural if no other sources are 
present, the soft masses  are all equal, with the squared mass equal to one 
third of the RHS in (\ref{genmass}). 
Then one gets $m_1^2=m_2^2=m_3^2=\ m^2$ with 
\beq
 m^2\ =\
\frac{g_s}{6}\left[ \vert G_{123} \vert^2 +
\frac{1}{4}\sum_{ij}(\vert S_{ij} \vert^2 + \vert A_{\bar{i}\bar{j}}
\vert^2) -
%\right. \\ & \left. -
Re(G_{123}G_{\bar{1}\bar{2}\bar{3}}+
\frac{1}{4}S_{lk}S_{\bar{l}\bar{k}}+\frac{1}{4}A_{lk}A_{\bar{l}\bar{k}})\right]
\label{redondo}
\eeq 
It is important to notice that in the presence of other sources not 
respecting the spherical symmetry, one in general generates additional 
mass terms which modify the scalar mass matrix (while preserving its 
trace).

\subsection{Examples}
\label{examples}

In this section we describe some prototypical examples illustrating a 
variety of possible classes of models. Clearly, combinations of the 
diverse ingredients can lead to interesting results as well.

\subsubsection{ISD fluxes and no-scale structure}

A familiar choice of flux configurations is to take the flux density 
$G_3$ to be imaginary self-dual, $*_6 G_3=iG_3$. This choice is 
interesting since in \cite{dgs,gkp} it was found that compactification on 
CY 
orientifolds with fluxes, to 4d Minkowski space, required the fluxes to 
satisfy this condition. In complex coordinates, the ISD condition allows 
for primitive $(2,1)$, non-primitive $(1,2)$ and $(0,3)$ components for 
$G_3$.

The explicit supergravity background describing ISD fluxes is of the form
\beqa
ds^2 & = & Z^{-1/2}\, \eta_{\mu\nu} dx^\mu dx^\nu\, +\, Z^{1/2}\, dx^m dx^m
\nonumber\\
\tau & = & \tau_0 \nonumber \\
G_3 & = & G_{lmn}\, dx^l dx^m dx^n = -i (*_6 G)_{lmn} \nonumber \\
\chi_4 & = & \frac{1}{Z}\, dx^0 dx^1 dx^2 dx^3 \\
F_5 & = & d\chi_4 + *_{10}d\chi_4 \nonumber 
\label{blackbrane}
\eeqa
with $Z= 1- (g_s/72)G^2 r^2$, with $G^2=G_{lmn}^* G^{lmn}$ and 
$r^2=(x^4)^2+\ldots (x^9)^2$ 
\footnote{Notice that we have chosen a spherically symmetric solution for 
the warp factor, etc, in order to fix the metric, dilaton and 5-form in 
terms of the fluxes.}.

As is obvious from the previous discussion, pure ISD fluxes do not give 
rise to any SUSY-breaking soft terms. Still, equations of motion 
only fix the trace of the mass$^2$ so that one has
$$
m_1^2+m_2^2+m_3^2\ =\ 0 
$$
If, as discussed above, one assumes spherically symmetric
boundary conditions one gets 
$$
m_1^2\ =\ m_2^2\ =\ m_3^2\ =\ 0 
$$
The absence of soft terms is a consequence of cancellations between the 
DBI and CS terms of the D3-brane action. Hence, they are consequence of 
the almost BPS nature of ISD fluxes \cite{gkp}, whose supergravity 
solution is of the black 3-brane form. Notice that this is not a 
consequence of supersymmetry, which is present only if the flux is 
primitive $(2,1)$ in some complex structure, while the imaginary 
self-duality condition in principle allows for additional components. 

The absence of soft terms is in agreement with the structure of the 
effective 4d supergravity of the compactifications in \cite{gkp}, which 
are of the no-scale type \cite{ns}, at leading order in $\alpha'$. This will be 
discussed in section \ref{sugra}.

\subsubsection{IASD fluxes}
\label{iasdfluxes}

Other interesting examples are  provided by D3-branes in an imaginary 
anti self-dual $G_3$ flux background, $*_6 G_3=-iG_3$. As discussed above, the 
configuration of D3-branes in IASD flux is equivalent to the situation of 
$\ov{\rm D3}$-branes in an ISD flux. Hence it provides the kind of 
soft terms on a $\ov{\rm D3}$-brane embedded in a compactification of the 
kind considered in \cite{gkp}. This is interesting, given the 
role played by such configurations in recent proposals to construct 
de Sitter vacua in string theory \cite{kklt}.

An explicit supergravity background with IASD flux  is given by eq.(3.10), 
except for an additional sign in the 5-form background, and the fact 
that $G_3$ is IASD. This implies that there are no cancelations in this case
\footnote{This has been discussed in \cite{verlinde} for the scalar 
trilinear couplings, which were argued to lead to D3-brane polarization 
for any IASD flux background. Our analysis corrects this point, by 
noticing that one in general also obtains non-trivial scalar masses, which 
prevent the scalars from acquiring vevs. Hence in general we obtain 
stable (or metastable) configurations of non-polarized D3-branes, even in 
the presence of IASD fluxes.}.
The IASD condition allows for $(3,0)$, primitive $(1,2)$ and non-primitive
$(2,1)$ components for $G_3$. Let us study them in turn.

\medskip

~\\

{\large {\it \bf a)  $(3,0)$ fluxes and dilaton domination}}

An interesting case is when the IASD flux has only 
$(3,0)$ component, i.e, only $G_{123}$ is non-vanishing.
From eq.(\ref{softgen}), (\ref{genmass2}), (\ref{redondo})
one gets (assuming spherically symmetric boundary conditions
for simplicity)
\beq
 m^2\ =\
\frac{g_s}{6}\, \vert G_{123} \vert^2 \ ;\ 
M^a\ =  \  {{g_s^{1/2}}\over {\sqrt{2}}}\ G_{123}
\ ;\ 
A^{ijk}\ =  \ -h^{ijk}\, {{g_s^{1/2}}\over {\sqrt{2}}}\ G_{123}
\label{dildom}
\eeq
with the rest of the soft terms vanishing. Note the relationships
\beq
A^{ijk}\ =\ - h^{ijk} M_a \ ;\ m^2\ =\ {1\over 3}|M|^2
\label{dildom2}
\eeq
This type of soft terms \footnote{This theory could be referred to as 
$\NN=0^*$ theory, in parallelism to the $\NN=1^*$ theory in \cite{ps}.} are 
remarkable 
in several respects and have appeared previously  in the literature.
Indeed they correspond to the {\it dilaton domination form } 
discussed in \cite{kl,bim}.
Namely, soft terms satisfying these relation appear in $\NN=1$ supergravity 
theories  with spontaneous supersymmetry breaking due to a vacuum 
expectation value for the auxiliary field in the dilaton multiplet.
The relation with our present configuration will be explained in 
section \ref{sugra}, where we show that in compactifications 
the $(3,0)$ component of the flux is indeed the auxiliary field of the 
dilaton (with respect to the supersymmetry unbroken by the D3-brane). 

Another interesting property of this type of boundary conditions,
noted a long time ago, is that their presence leads to a 
$\NN=0$ theory which is finite in all orders in perturbation theory
\cite{finite1,finite2}.
We will come to this interesting property in section (\ref{sugra}).

~\\

{\large {\it \bf b)  $(1,2)_P$ fluxes and `susy' soft terms}}

Let us now consider configurations of a pure primitive (1,2) flux, which 
we parametrize in terms of the tensor $S_{ij}$. From (\ref{softgen}), 
(\ref{genmass2} ), one gets 
\beq
m_1^2 + m_2^2 + m_3^2\, =\,
\frac{g_s}{8}\, \sum_{ij}\, \vert S_{ij} \vert^2  \ ;\ C^{ijk}\,  =\,   
-h^{ijl}\, \mu _{kl} \ ;\ \mu _{ij}\,   =\,   -{{g_s^{1/2}}\over 
{2\sqrt{2}}}\, S_{ij}
\label{12psoft}
\eeq
with the rest of the terms vanishing. It is easy to convince oneself that 
these terms are consistent with an effective $\NN=1$ SUSY theory 
corresponding to a superpotential
\beq
W \ =\ W_{\NN=4}\ +\ \frac {1}{2}  \mu_{ij}\Phi^i \Phi^j
\label{12superpot}
\eeq
Consider for example the case in which only $S_{11}$ is non-vanishing.
The above results are consistent with an $\NN=1$ theory with 
two massless chiral multiplets, $\Phi^{2,3}$, and one massive multiplet
$\Phi^1$ with mass ${{g_s^{1/2}}\over {2\sqrt{2}}}\, S_{11}$.
It is also easy to check that fluxes with just one off-diagonal component
$S_{ij}$ lead to terms consistent with an $\NN=2$ SUSY.

This type of fluxes are of particular interest since they give rise to 
masses for non-chiral fermions. Such terms are required for instance in 
order to obtain a `$\mu $-term' for the Higgs multiplets  in the MSSM.

The reason underlying the $\NN=1$ structure of soft terms for this kind 
of fluxes is presumably their relation to the flux background in the 
Polchinski-Strassler construction \cite{ps}. Indeed, it is possible to 
complement the above primitive $(1,2)$ fluxes with a suitable combination 
of a $(0,3)$ component, and a non-primitive $(1,2)$ component, such that 
they obey the constraints (3.30) in \cite{gp}. These additional pieces do 
not modify the soft terms, because they are ISD components (so that they 
do not generate trilinear terms or fermion masses) and such that 
the `interference term' in scalar masses (\ref{genmass2}) vanishes. The 
conditions in 
\cite{gp}, satisfied by this completed flux, guarantee that the system of 
the D3-brane and the flux preserve an $\NN=1$ supersymmetry (although it 
is not the same preserved by a pure D3-brane system) and that the 
deformation of the $\NN=4$ theory is provided by a 
superpotential $W$ related to the original $(1,2)_P$ flux as obtained 
above by direct analysis, namely $S_{ij}\sim \partial_i\partial_j W$
\footnote{An important point is that in our analysis we expand the 
worldvolume theory of a D3-brane, whereas in \cite{ps} the 
supersymmetric ground state was shown to correspond to dielectric 
D3-branes. Hence both D3-brane systems are different (meta)stable 
configurations of the same underlying $\NN=1$ supersymmetric lagrangian.}.
It would be interesting to understand the relation to other 
supersymmetric flux configuration in the literature \cite{susyflux}.

~\\

{\large {\it \bf c)  $(2,1)_{NP}$ fluxes}}

This flux is parametrized in terms of $A_{{\bar i}{\bar j}}$.
From eq.(\ref{softgen}), (\ref{genmass2}) one gets
\beq
m_1^2 + m_2^2 + m_3^2 \, =\,
\frac{g_s}{8}\, \sum_{ij}\, \vert A_{{\bar i}{\bar j}} \vert^2  \ ;\ 
C^{ijk} \, =\, -h^{ijl}\, \frac {g_s^{1/2}}{\sqrt{2}}\, 
(A_{{\bar l}{\bar k}})^*\ ;\  M_g^{ia}\, =\, {{g_s^{1/2}}\over 
{4\sqrt{2}}} \, \epsilon_{ijk} \, A_{{\bar j}{\bar k}}
\label{sosos}
\eeq
These soft terms break supersymmetry completely. 
In CY compactifications, such type of flux cannot be turned on (since the 
existence of non-primitive fluxes requires the existence of non-trivial 
5-forms to support the non-vanishing $G_3\wedge J$), so this situation
is less interesting.

\subsubsection{Mixed ISD and IASD fluxes}
\label{gensit}

It is interesting to consider more general situations, with fluxes not 
purely ISD or IASD. For dim=3 soft terms both ISD and IASD 
contributions add linearly. However, as observed from
eq.(\ref{genmass}), the contributions to scalar masses 
upon imposing equations of motion do not add linearly, there
is an interference term proportional to the products of
ISD and IASD contributions. In particular if one has
\beq
G_{{\bar 1}{\bar 2}{\bar 3}}= G_{123}\ ;\
S_{{\bar i}{\bar j}}=S_{ij} \ ;\ A_{ij}=A_{{\bar i}{\bar j}}
\label{toseva}
\eeq
one can check that the 5-form background vanishes and 
the scalar masses also do. More general situations are obtained 
for arbitrary ISD and IASD fluxes.

As a prototype class of models, we consider turning on 
a flux with $(3,0)$ and $(0,3)$ components, namely
\beqa
G_3 & = & G_0\, dz_1\, dz_2\, dz_3\, +\, G_0{}'\, d{\ov z}_1\, d{\ov 
z}_2\, d{\ov z}_3.
\eeqa
In this case the supergravity solution is not of the black 3-brane form, 
so we  use our general ansatz (\ref{ansatz}) and impose the 
equations of motion (\ref{eomcomplete}) on the coefficients of 
(\ref{powerexp1}). We obtain
for the soft terms (and spherically symmetric boundary conditions)
\beqa
m^2\ & =&\ \frac{g_s}{6}\, \left(\, \vert G_{0} \vert^2  -
\preal (G_{0}G_0{}')\, \right) \nonumber \\
M^a\ &= & \  {{g_s^{1/2}}\over {\sqrt{2}}}\ G_{0}
\ ;\
A^{ijk}\ =  \ -h^{ijk}\, {{g_s^{1/2}}\over {\sqrt{2}}}\ G_{0}
\label{mixto}
\eeqa
Clearly, allowing for these more general fluxes one is able to reproduce 
a rich pattern of soft terms on the D3-brane worldvolume. In 
particular, without loss of generality one can parametrize, (assuming 
$G_0\not =0$, which would just give rise to no soft terms)
\beq
G_{123} \ =\ |G_0|\, e^{i\delta} \ \ ;\ \ G_{\bar{1}\bar{2}\bar{3}}\ =\ 
\tan\theta\, e^{i\beta}\, 
\vert G_0 \vert \ 
\eeq
with $\delta$, $\beta$ real parameters. In terms of these  one obtains for 
the scalar masses
\beqa
m^2\ &=&\ { {|M|^2}\over 3}\, [\, 1-\tan\theta\, \ cos(\delta +\beta )\, ] 
\nonumber \\ 
A^{ijk}\ &=&\ -M h^{ijk}  \ \ 
\eeqa
Note that this more general situation alters only the relative size of 
the scalar masses, for which a variety of possibilities now appear.  
Consider for example real flux backgrounds ($\beta =\delta =0$ mod 
$2\pi$). For $\tan\theta =0$ one recovers the dilaton dominated soft terms 
discussed in the previous section. For $\tan\theta =1$ one gets vanishing 
scalar masses but non-vanishing $A$ and $M$ parameters. This corresponds 
to the case discussed above with a vanishing 5-form. Note that in this case 
the trilinear couplings will destabilize the branes since a non-vanishing 
vev for the scalars will be energetically preferred. Finally, for 
$\tan\theta \ll -1$ one gets $m^2\gg M^2$ and scalar masses are much 
greater 
than gaugino and trilinear terms.

The above expressions could be interesting for phenomenological SUSY 
analysis if applied to the more realistic theories (yielding an open 
string spectrum close to the MSSM) instead of a $\NN=4$ theory. Some 
steps in this direction are taken in sections \ref{orbifolds}, 
\ref{realistic} .

\section{SUSY-breaking by distant antibranes}
\label{antibranes}

Our general formalism to compute the SUSY-breaking soft terms is not 
restricted to configurations of fluxes. Rather, it allows the computation 
of soft terms in other non-supersymmetric situations, even if no fluxes 
are present. In particular, it provides an interesting way to compute the 
effect of supersymmetry breaking due to a set of distant antibranes, and 
can also be applied to compactifications based on \cite{adsau}, e.g. with 
a visible sector of supersymmetric D-branes, and supersymmetry broken in 
a sector of antibranes. This is interesting since there exist models with 
semirealistic spectrum in this class \cite{aiq,aiqu}, so it is natural to 
explore the basic features of SUSY-breaking soft terms in this setup.
Moreover, anti-D3-branes play a prominent role in recent descriptions of 
de Sitter vacua in string theory \cite{kklt}. 

Some relevant observations of this approach are discussed in this 
section, and we leave a more systematic discussion for future work.
For the purposes of a general discussion, we center on the soft terms 
induced on a D3-brane (at the origin in $\IR^6$) due to the background 
created by a stack of $M$ distant anti-D3-branes (at the point $\vec{x}_0$
in $\IR^6$). In order for the supergravity background at 
the D3-brane 
location to be trustable, one should be in the regime where the 
$\ov{\rm D3}$-branes are distant. Luckily, this is also the interesting 
situation, where one is far from the tachyonic regime, and also the 
situation of interest in brane inflation discussions \cite{inflation}. 

The background is given by
\beqa
ds^2 & = & Z(x^m)^{-1/2}\, \eta_{\mu\nu} dx^\mu dx^\nu\, +\, 
Z(x^m)^{1/2} \, dx^mdx^m \nonumber\\
\chi_4 & = & \frac{1}{Z}\, dx^0 dx^1 dx^2 dx^3 \nonumber\\
F_5 & = & d\chi_4 + *d\chi_4 \nonumber\\
Z & = & 1-\frac{1}{2\pi^2}\, \frac{M}{|\vec{x}-\vec{x_0}|^4}
\eeqa
and there is a vanishing NSNS and RR flux background.

The computation of some terms from the DBI action is described in appendix 
\ref{moregen}. Clearly, the leading terms show the instability of the 
configuration, namely a linear term and tachyonic mass terms for the 
worldvolume scalars, signalling the tendency of the D3-branes to approach 
the anti-D3-brane stack. Such instabilities can be avoided in explicit 
examples (like branes stuck at singularities,
see section (\ref{realistic})), which are more involved but 
presumably retain the other main features of this simpler configuration. 
Hence we proceed to discuss these aspects, ignoring the 
instability-related terms. Instead of entering the detailed discussion, we 
prefer to make some general comments on soft terms due to antibrane 
supersymmetry breaking.

Antibrane supergravity backgrounds contain basically a warp factor and 
a 5-form background, and do not introduce any 3-form fluxes. Therefore, 
at the order we have studied, the soft terms on D3-branes would be just 
scalar masses (and possibly trilinear terms, arising from the metric and 
5-form). In sharp contrast with the above examples with fluxes, the 
present situation does not lead to fermion masses.

This is strongly suggestive that the supersymmetry breaking induced by 
antiD3-branes is a D-term breaking. Indeed this nicely dovetails with 
properties of D-branes on Calabi-Yau manifolds, where the tension (or 
vacuum energy) associated to B-type branes (namely, branes wrapped on 
holomorphic cycles; in our case of $\ov{\rm D3}$-branes at a point in the 
Calabi-Yau) is controlled by K\"ahler parameters, which couple to 
D3-branes 
as D-terms \footnote{In this respect, in \cite{quevedo} stacks of 
anti-D3-branes were replaced by D7-brane worldvolume anti-instantons, 
whose tension couples as D-terms.}. It would be nice to derive this from 
a 4d supergravity approach, as done for flux SUSY-breaking in section
\ref{sugra}.

From the phenomenological viewpoint, the absence of gaugino masses is an 
important drawback for this kind of supersymmetry breaking.
Concerning the scalar potential, it is difficult to make model-independent 
statements. However, a general statement is that, since the anti-D3-branes 
are distant, they give no additional sources at the D3-brane location
and the relation (\ref{genmass}) is unchanged (and similarly for the 
5-form and dilaton equations). Hence the addition of distant antibranes in 
general changes the scalar masses, but leaves the trace of the scalar mass 
matrix unchanged (see section \ref{massesps} for further discussion and an 
explicit example). Hence, in models where the only source of scalar masses
are distant antibranes, at least some scalars are tachyonic \footnote{In 
a sense, this is related to the instability issues: for instance, 
D-branes stuck at singularities avoid this problem because they may not 
contain the corresponding scalars.}.

In conclusion, although very interesting from the theoretical viewpoint, 
this source of SUSY-breaking, taken by itself, would not be 
phenomenologically viable. However, if distant antibranes are present in a 
configuration with other structures, like 3-form fluxes, they may lead to 
interesting substructures in the pattern of soft terms, like deviations from 
universality in an otherwise dilaton-domination SUSY-breaking 
configuration. This departure from universality, if applied to
a realistic model, lead in general to Flavour Changing
Neutral Currents (FCNC). These are also very much constrained 
experimentally,  so distant antibrane SUSY-breaking 
should again not be the dominant source in order to
avoid further phenomenological problems.
We discuss in certain detail a semirealistic example 
in section \ref{realistic}.

\section{D7-branes and fluxes}
\label{dseven}

It is interesting to point out that the absence of worldvolume soft terms
in the presence of ISD fluxes is a special feature of D3-branes, and does
not hold for other D-branes. For instance, it has been argued \cite{tt} that
D7-branes in compactifications of the kind in \cite{gkp} with ISD 3-form
fluxes, have their moduli stabilized by the ISD flux. That is, the flux
induces non-trivial mass terms for the D7-brane worldvolume scalars.

Clearly one can apply our general techniques to analyze the
D7-brane worldvolume soft terms, again by simply coupling their DBI+CS
action to the supergravity background. We leave this general problem for
future research, but here provide the main features of soft terms for
the interesting case of D7-branes in flat space, in the presence of ISD
fluxes.

The effect of the background (3.10) on a D7-brane probe is
quite simple. Indeed, since a D7-D3 system is BPS, and the ISD flux
mimics the dilaton, metric and 5-form background of a D3-brane, it is
clear that the dilaton, metric and 5-form in (\ref{blackbrane}) lead to no
soft terms on the D7-brane worldvolume. This follows from cancellations
between diverse pieces in its action. 

The only difference between an ISD and a D3-brane background is the 
presence of $G_3$ itself in the former. Hence, the soft terms we expect 
on D7-branes in ISD fluxes are those soft terms generated by $G_3$ 
directly, without use of the supergravity equations of motion. For 
instance, one immediate possible effect is the appearance of fermion 
masses, and scalar trilinear terms, in analogy with the above. 

In addition and interestingly enough, there exists a new kind of soft 
term, namely scalar masses directly induced by the 3-form fluxes. The new 
contribution arises from the D7-brane DBI determinant, the analog of 
the last line in (\ref{determ}), involving the NSNS 2-form field in
worldvolume directions transverse to 4d Minkowski space (denoted
$a,b=4,5,6,7$)
\beqa
\, [ -\det ( P [E_{\mu\nu}] ) ]^{1/2} & = & 1 -
\frac{\sigma^2}{2}\, \partial_\mu \phi^m\,
\partial^\mu \phi^m \,+\, \sigma^2\,  K_{mn}\, \phi^m\phi^n
-\frac{1}{2} B_{ab} B^{ab}
\eeqa
Using that $B_{ab}=\frac{1}{3}H_{abr} x^r+\ldots$, with $r=8,9$, and 
trading $x^r$ for
$\phi^r$, we obtain additional soft terms involving masses for $\phi^r$
\beqa
\frac{1}{18}H_{abr}H^{ab}~_s \phi^r \phi^s
\eeqa
Notice that there are no additional terms of this kind from the CS action, 
hence this
contribution does not cancel out, and provides the non-zero scalar masses
for D-brane worldvolume scalars, even for ISD fluxes.

The spacetime interpretation of this result is that if D7-branes move in
transverse space, an induced D5-brane charge appears due to the
non-trivial $B_{NS}$ field. This increases their tension, hence the
D7-brane motion is energetically costly \footnote{We thank S. Trivedi and
P. Tripathy for suggesting this mechanism.}. The above scalar mass term
accounts precisely for this additional tension, as is obvious from its
DBI origin.

A more general argument, described in the literature, exploits the 
description of CY threefold compactifications with 3-form fluxes and 
7-branes in terms of F-theory (of M-theory in a dual picture) on an 
elliptic CY fourfold with 4-form fluxes. In the latter, 7-branes are 
encoded in complex structure parameters, describing the geometry of the 
degenerations of the elliptic fibration, and which are fixed by the 4-form 
fluxes. Hence one concludes that 7-brane moduli are fixed by 3-form 
fluxes.

Notice that the existence of mass terms for D7-brane worldvolume 
fields for ISD fluxes does not spoil the no-scale structure of the 
resulting 4d effective theory. Indeed, the masses of the D7-brane moduli 
are given by the flux density scale, hence is of the same order as masses 
of the complex structure moduli. This is also clear in the F-theory 
picture, where both kinds of moduli are on an equal footing.

\section{D3-branes at orbifolds}
\label{orbifolds}

Our results so far have been concerned with configurations of D3-branes
in flat space, or at a smooth point in the transverse space. Consequently
the D3-brane worldvolume field theory is a deformation of the $\NN=4$
theory, and hence is non-chiral. 

One of the natural ways to apply our results to the phenomenologically 
interesting setup of soft terms in chiral supersymmetric gauge sectors, 
would be to 
consider the D3-branes to sit at a singular point in transverse space. 
Moreover, one of the simplest choices of singularities, namely orbifold 
singularities, are just quotients of flat space by a discrete subgroup 
$\Gamma$ of the $SO(6)$ rotational isometry group in transverse space. 
Correspondingly the D3-brane worldvolume theory (denoted orbifold gauge 
field theory for short) is obtained from the underlying $\NN=4$ $U(N)$ 
theory by projecting out fields non-invariant under such action \cite{dm}. 
In this projection, one should take into account (on top of the action on 
Chan-Paton (CP) labels) that the gauge bosons are singlets under the 
$SO(6)$, while scalars and fermions transform in the representations 
${\bf 6}$, ${\bf 4}$.

The introduction of fluxes in the setup of D3-branes at singularities is 
straightforward, by regarding the configuration in the covering space. In 
the latter,
we have D3-branes in a smooth supergravity background, on which only 
$\Gamma$-invariant supergravity fields are allowed. On the covering space, 
they induce 
a set of soft terms for the fields in the underlying parent $\NN=4$ theory, 
as above. Due to the $SO(6)$ invariance of the above soft terms, the 
$\Gamma$-invariant supergravity background induces soft terms for 
$\Gamma$-invariant worldvolume fields, i.e. fields which descend to 
the orbifold field theory. Now going to the quotient, this implies that 
the supergravity background around the orbifold singularity induces the 
above soft terms in the orbifold gauge field theory. Notice that $SO(6)$ 
invariance is a crucial ingredient, guaranteeing that certain inconsistent
soft terms (like mass terms for the chiral fermions present after the 
orbifold projection) are not present because the corresponding flux is 
projected out by the orbifold.

A particularly simple case is that of Abelian orbifold $\IC^3/\IZ_N$ 
singularities. The twist is given by a generator $\theta $ acting like
\beqa
\theta:(z_1,z_2,z_3) \to (e^{i2\pi a_1/N} z_1, e^{i2\pi a_2/N} z_2,
e^{-i2\pi a_3/N} z_3)
\eeqa
which leads to one unbroken SUSY on the D3-brane worldvolume (before 
adding fluxes) if $a_1+a_2+a_3=0$, mod $N$. If all $a_i\not= N/2$, only
fluxes of types $(0,3)$ and $(3,0)$ will be invariant under the twist. On 
the other hand, if one has e.g., $a_3=N/2$ some primitive $(2,1)$ and 
$(1,2)$ fluxes will be consistent with the twist. 

These features are more concretely discussed in explicit examples. Consider
the $\IC^3/\IZ_3$ singularity, where the generator $\theta$ of $\IZ_3$ acts
as 
\beqa
\theta:(z_1,z_2,z_3) \to (e^{2\pi i/3} z_1, e^{2\pi i/3} z_2, 
e^{-4\pi i/3} z_3)
\eeqa
Before the introduction of fluxes, the orbifold gauge field theory has 
the following $\NN=1$ multiplet content
\begin{center}
\begin{tabular}{ccc}
$V$ & Vect. Mult. & $ U(n)\times U(n)\times U(n)$ \\
$\Phi^1_{a,a+1}$ & Ch. Mult. & $(n,\bar{n},1) + (1,n,\bar{n}) + 
(\bar{n},1,n)$ \\
$\Phi^2_{a,a+1}$ & Ch. Mult. & $(n,\bar{n},1) + (1,n,\bar{n}) + 
(\bar{n},1,n)$ \\
$\Phi^3_{a,a+1}$ & Ch. Mult. & $(n,\bar{n},1) + (1,n,\bar{n}) + 
(\bar{n},1,n)$
\end{tabular}
\end{center} 
The allowed fields in the supergravity background are $ K_{i\bar{\j}}$, 
$\tau_{i\bar{\j}}$, $ \chi_{i\bar{\j}}$, and $G_{123}$, 
$G_{\bar{1}\bar{2}\bar{3}}$, denoted $G_0$, $G_0{}'$ respectively in what 
follows.

So, due to the orbifold projection, the most general supergravity 
background is of the form considered in section \ref{gensit}. The soft 
terms are therefore
\beqa
{\cal L} & = &\Tr \biggr[ \,
-\, (\, 2K_{i\bar{\j}}- \chi_{i\bar{\j}}+g_s (\pim 
\tau)_{i\bar{\j}}\,)
\, \Phi^i\Phi^{\bar{\j}}\,   \nonumber\\
&& +\, \frac{g_s\sqrt{2\pi}}{3}\, G_0\, \epsilon_{ijk}\, 
\Phi^i\Phi^j\Phi^k\, +\, {\rm  h.c.} \, 
+\, \frac{g_s^{1/2}} {2\sqrt{2}} 
\, G_0\, \lambda\lambda\, +\, {\rm  h.c.} \biggr]
\eeqa  
where it is implicit that only gauge invariant combinations are present, 
namely $\Phi^i\Phi^{\bar{\j}}$ stands for $\sum_a \Phi^i_{a,a+1} 
(\Phi^j_{a,a+1})^*$

Notice that in particular no mass terms for the fermions in the chiral 
multiplets have been generated. Hence we see a nice interplay of the 
orbifold projection in the worldvolume field theory (where it leads to 
chirality via projection from the parent $\NN=4$ theory) and in the 
supergravity background (where it forbids a background for non-invariant 
fields), correlated via the $SO(6)$ invariance of the underlying theory.

\smallskip

The above discussion can be carried out similarly for other orbifolds. It 
is interesting to notice that when the orbifold field theory contains 
non-chiral fermions, the fluxes that generate soft masses for them are 
invariant under the orbifold action. For instance, for orbifolds 
$\IC^2/\IZ_N\times \IC$, with the $\IZ_N$ generated by 
$\theta:(z_1,z_2,z_3)\to (e^{2\pi i/N} z_1, e^{-2\pi i/N} z_2, z_3)$,
the worldvolume theory has $\NN=2$ before the introduction of fluxes. 
Hence, the $\NN=1$ chiral multiplet $\Phi_3$ contains adjoint fermions, 
$\Psi_3$, which are non-chiral. Correspondingly, one can consider 
$\IZ_N$-invariant IASD fluxes of the form 
\beqa
G_3\, =\, G_{123}\, dz_1 dz_2 dz_3\,+\, G_{\bar{1}\bar{2}3}\,
 d{\bar{z}}_1 d{\bar{z}}_2 dz_3
\eeqa
which generically introduce soft terms breaking the $\NN=2$ D3-brane 
worldvolume supersymmetry; in particular, the second term introduces 
a soft mass for $\Psi_3$. 

\smallskip

An interesting example, leading to a chiral spectrum which contains a 
non-chiral subsector (which can thus acquire mass terms) is provided by 
the $\IC^3/\IZ_4$ orbifold, generated by $\theta:(z_1,z_2,z_3)\to (e^{2\pi 
i/4} z_1, e^{2\pi i/4} z_2, -z_3)$.

The spectrum of $\NN=1$ multiplets is given by \footnote{We consider 
D3-branes in the regular representation of $\IZ_4$, although slightly 
more general choices can be made, consistently with twisted RR tadpole 
cancellations.}
\begin{center}
\begin{tabular}{ccc}
$V$ & Vect. Mult. & $ U(n)_1\times U(n)_2\times U(n)_3\times U(n)_4$ \\
$\Phi^{1,2}_{a,a+1}$ & Ch. Mult. & $2\, [\, (\bar{n}_1,n_2) + 
(\bar{n}_2,n_3)+ (\bar{n}_3,n_4) + (\bar{n}_4,n_1)$ \\
$\Phi^3_{a,a+2}$ & Ch. Mult. & $(n_1,\bar{n}_3)+ 
(\bar{n}_1,n_3) + (n_2,\bar{n}_4) + (\bar{n}_2,n_4)$ \\
\end{tabular}
\end{center} 
The allowed non-trivial supergravity background parameters are 
$K_{i\bar{j}}$, for $i,j=1,2$, $K_{33}$, $K_{\bar{3}\bar{3}}$, 
$K_{3\bar{3}}$ (and similar ones for $\tau_{mn}$ and $\xi_{mn}$), and 
$G_{123}$, $G_{\bar{1}\bar{2}\bar{3}}$, $G_{12\bar{3}}$, 
$G_{\bar{1}\bar{2}3}$. Thus, the most general allowed soft term 
lagrangian is
\beqa
{\cal L} && = \Tr \biggr( \,
-\, \sum_{ij=1,2} (\, 2K_{i\bar{\j}}- \chi_{i\bar{\j}}+g_s (\pim 
\tau)_{i\bar{\j}}\,) \, \Phi^i\Phi^{\bar{\j}}\,  -  
\, \frac{g_s^{1/2}}{2\sqrt{2}}\, (\, G_{123}\, \lambda\lambda\, 
+\,  \frac {1}{2}\, S_{33}\, \Psi^3\Psi^3) \, +\, {\rm  h.c.}\, 
 \nonumber\\
&& - \, \frac{1}{2}(\, 2K_{33}+ \chi_{33}+g_s (\pim \tau)_{33}\,) \, 
\Phi^3\Phi^{3}\, +\, {\rm h.c.}\, - \, (\, 2K_{3\bar{3}}- 
\chi_{3\bar{3}}+g_s (\pim \tau)_{3\bar{3}}\,) 
\, \Phi^3\Phi^{\bar{3}}\,   \nonumber\\
&& +\frac{g_s\sqrt{2\pi}}{3}\,  \left[\, G_{123}\, \epsilon_{ijk} \,
\Phi^i\Phi^j\Phi^k \, +\, {3\over 2}\,
\epsilon_{\bar{i}\bar{j}\bar{3}}\,  S_{33} \,
\Phi^{\bar{i}}\Phi^{\bar{j}}\Phi^3 \, \right]\,  +\,
\, {\rm  h.c.}  \biggr) 
%&& +
\eeqa

where again it is implicit that only gauge invariant terms appear, namely
$\Phi^3\Phi^{3}$ denotes couplings $\Phi_{1,3}^3\Phi^{3}_{3,1}+ 
\Phi^3_{2,4}\Phi^{3}_{4,2}$ (and similarly for $\Psi^3\Psi^3$).

As we already mentioned, one interesting feature of this example is that 
the orbifold symmetry is compatible with flux components that lead to 
$\mu $-terms and 
B-terms for some worldvolume scalars. Indeed one has a $\mu $-term
for the $\Psi^3$ fermion $\mu _{33} = g_s^{1/2}S_{33}/(2\sqrt{2})$. 
The B-terms arise arise from the coefficients 
$K_{33}$, $\tau_{33}$, $\chi_{33}$ for the 10d metric, dilaton and 5-form 
background
\footnote{Note that, as emphasized, this kind of soft terms
will arise in the worldvolume of anti-$D3$-branes in the 
presence of ISD $(0,3)$ and $(2,1)_P$ fluxes also.
In particular, SUSY-preserving $(2,1)_P$ fluxes give rise
to no soft terms on the worldvolume of $D3$-branes 
but give rise to a '$\mu $-term' on that of antibranes.} . It is 
interesting to point out that B-terms do not contribute to the trace of 
the mass matrix, and therefore are completely unrelated to the flux 
background, and depend crucially on the boundary conditions for fields
in our local configuration (i.e. depend on the structure of 
distant sources or the global properties of the compactification). It is 
tempting to speculate that this may lead to interesting mechanisms to 
generate phenomenologically viable values for the B-term in MSSM-like 
models.

\smallskip

Finally we would like to point out that modding out the above $\IZ_3$ 
theory by the group of permutations of the three complex planes leads to 
a configuration of D3-branes at the $\Delta_{27}$ singularity
\cite{aiqu,jei}. One 
immediately recovers the background, field theory and soft terms 
reproducing \footnote{Our results slightly correct those 
in \cite{grana} in a few numerical coefficients.} those in 
\cite{grana}. Our approach in this paper is more general, and allows to 
study D3-branes at smooth points, and at other orbifolds. It would be 
interesting to device a way of computing soft terms for D3-branes at 
other interesting, but non-orbifold, singularities, like conifolds and 
generalizations thereof.

\section{SUSY-breaking soft terms and the effective $\NN=1$
supergravity effective action }
\label{sugra}

The approach to compute soft terms we followed up to now was 
very general in the sense that we were able to compute them 
in terms of the {\it local } Type IIB supergravity
backgrounds. In general we would be interested in full-fledged 
compact vacuum solutions, but in order to compute soft terms in the
worldvolume of D3-branes that local information was enough.

In a full compactification  we would like to interpret  
those terms as arising from 
vevs for the auxiliary components  of some 4d chiral superfields.  
In a general CY compactification the natural candidates for
such chiral fields will be the dilaton $S$ and K\"ahler and 
complex structure moduli.
 A simple class of models are the type IIB Calabi-Yau 
orientifold models in \cite{dgs,gkp} (which include the explicit 
${\IT}^6/\Omega R$ models in \cite{fp,kst,kors}, and others \cite{tt}). 
These authors showed that the presence of RR and NSNS fluxes is able to 
fix the values of the dilaton and generically all complex structure 
moduli, while K\"ahler moduli are in general not fixed. Our analysis in 
this section will apply
to any such CY orientifold in the large radius limit, in which
the behaviour of the K\"ahler potential is quite general.

\subsection{SUSY-breaking soft terms and $\NN=1$ auxiliary fields}

In a general CY (not $\IT^6$ or K3$\times \IT^2$) there are no 
non-trivial 5-forms. This implies that it is not possible to turn on 
non-primitive fluxes, since $G_3\wedge J\neq 0$ would define a non-trivial 
5-form. Our analysis will center on general CY compactifications, hence 
the ISD condition will allow for $(0,3)$ and primitive $(2,1)$ pieces, 
and the IASD condition will allow for $(3,0)$ and primitive $(1,2)$ 
pieces. The analysis for $\IT^6$ or 
K3$\times \IT^2$ is complicated due to the presence of additional 
supersymmetry and hence additional auxiliary fields. See 
\cite{ferrara,kors} 
for some discussion of $\IT^6$ compactification.

The massless spectrum in any CY-orientifold always includes the complex 
dilaton chiral field
\beq
S\ =\ -i\tau \ \quad ;\ \quad \tau \ =\  C\ +\ i/g_s  \ ,
\label{sfield}
\eeq
whose imaginary part is related to the dilaton and  $C$ is the type IIB 
axion. In addition there will be a number of complex structure   
scalars $M_a$ and K\"ahler moduli $T_\alpha$ (18+18 in the $T^6/\Omega R$ 
example of \cite{fp,kst}), characterizing the compact volume. In order to 
simplify matters we will focus first  on the dynamics of the dilaton chiral 
field $S$ and the overall K\"ahler modulus $T$, whose real part gives the 
overall radius  of the compactification. More precisely (see e.g.   
 refs.\cite{gkp,red,kors}), we will have $T=\exp{4\sigma} +ib$,
with $\exp{\sigma}$ being the breathing mode of the compact
manifold volume. In the large volume 
limit the K\"ahler potential has the well known form
\beq
{K\over {M_p^2}} \ = \ -\log(S+S^*)\ -\ 3\log(T+T^*)
\label{kapot}
\eeq
where $M_p^2$  is the 4d Planck mass squared. We would now like to use the 
effective lagrangian in order to reproduce some of the results obtained 
for soft terms in the previous sections.  We consider, as in 
\cite{dgs,gkp,fp,kst}, compactifications with a non-trivial background for
the NSNS and RR field strength 3-forms $H_{(3)}$,$F_{(3)}$. In this 
situation an $S$-dependent effective superpotential is  given by
\cite{superp}
\beq
W\ =\  \kappa_{10}^{-2}
\int G_{(3)}\wedge \Omega  \ \ ,\ \ G_{(3)}\ =\ 
F_{(3)}-iSH_{(3)} \ ,   
\label{superp}
\eeq
where $\kappa _{10}^2={1\over 2}(2\pi )^7\alpha '^4$ is the 
$D=10$ gravitational constant and
 $\Omega$ the Calabi-Yau holomorphic 3-form. 

The supergravity auxiliary field  for a chiral field with 
K\"ahler metric $K_{i\bar{\j}}$ is given by
\beq 
{\overline F}^{\bar i}\, =\, \exp(K/2M_p^2)\,\,K^{\bar{i}j}\ D_j W/M_p^2 
\label{aux}
\eeq
with indices meaning derivation, and $D_jW=\partial_jW+K_j W$. 
Applying  this to  the $S$ and $T$ fields, one finds using 
the above expression for the superpotential,
\beqa
F^S\ &=\ {1 \over 
{M_p^2}}\, (S+S^*)^{1/2}\, (T+T^*)^{-3/2}\, 
(\kappa_{10}^{-2})\int G_{(3)}{}^*\wedge \Omega 
\nonumber \\ 
F^T\ &=\ -{1 \over {M_p^2}}\, 
(S+S^*)^{-1/2}\, (T+T^*)^{-1/2}\, (\kappa_{10}^{-2})\int G_{(3)}\wedge \Omega
\eeqa
Note that these equations show that a flux background of $(3,0)$ type 
would then correspond to a non-vanishing auxiliary field for  the complex 
dilaton $S$, whereas a $(0,3)$ background  would correspond to the
overall K\"ahler field $T$. One also finds for the gravitino mass
\beq
m_{3/2}^2 \ =\  { {e^{K/M_p^2} }\over {M_p^4}}\, |W|^2\ = \  
{1 \over {M_p^4}}\, (S+S^*)^{-1} \,
(T+T^*)^{-3}\, (\kappa_{10}^{-4})\, |\int G_{(3)}\wedge \Omega\,|^2 \ \ .
\label{gravit}
\eeq
One can then compute the vacuum energy in terms of the auxiliary fields 
to be given  by  
\beqa
V_0\ &=&\ M_p^2\, (\, F^{S^*}K_{SS^*}F^S\, +\, F^{T^*}K_{TT^*}F^T\, )\,
-\, 3\, m_{3/2}^2\, M_p^2 \ =\ M_p^2\, F^{S^*}K_{SS^*}F^S\ =
\nonumber \\ 
\ &=&  { {\kappa_{10}^{-4}|\int G_{(3)}^*\wedge \Omega|^2} \over 
{M_p^2  (S+S^*)(T+T^*)^3} } \ \ .
\label{vnot}
\eeqa
As expected, the contribution of the $T$ auxiliary field to the vacuum 
energy precisely cancels the $-3m_{3/2}^2$ supergravity term. This is 
just the well known cancellation taking place in no-scale models 
\footnote{This structure presumably does not survive $\alpha'$ and 
$g_s$ corrections (see \cite{bbhl} for a discussion of some $\alpha'$ 
correction terms). \label{alphaprime}}. Thus only the complex dilaton 
auxiliary field contributes to 
the vacuum energy. Note that if one minimizes the energy, as would be 
required by the equations of motion, the dynamics of the system yield 
$\int G_{(3)}^*\wedge \Omega=0$, so that the $(3,0)$ piece vanishes, in 
agreement with the results in \cite{gkp}. (In order to eliminate other 
IASD components, minimization with respect to complex structure moduli 
should be imposed as well, see below).
 
\medskip

In addition to the dilaton and moduli fields, these compactifications 
include gauge sectors on D3-branes, which generically sit at smooth points 
(or less generically at a singular point)
in the internal space.  We would like to analyze, using 
the effective lagrangian approach, the structure of the SUSY-breaking soft 
terms induced by the fluxes. For such purpose we can use standard $\NN=1$ 
supergravity formulae, already employed in the past to study 
soft terms in heterotic 
vacua (see ref.\cite{bimrev} for a review). 

Masses for the gauginos of the $a^{th}$ gauge factor may be obtained from 
the general formula
\beq
M_a \ =\ (2\preal f_a)^{-1} F^i \ \partial_i f_a
\label{gaugmas}
\eeq 
where $f_a$ is the gauge kinetic function. In the present D3-brane case
this gauge kinetic function is simply given by $f_a=S$, so that one easily 
finds for gaugino masses
\beq
M_a \ =\  {1 \over 
{M_p^2}}(S+S^*)^{-1/2}(T+T^*)^{-3/2}\ (\kappa_{10}^{-2})\int G_3^*\wedge \Omega
\label{gaugmas2}
\eeq
One can also use general expressions to compute soft terms  
trilinear in the adjoint scalars $\Phi_i$. For this we need to know the
K\"ahler potential $K$ of the scalar fields. In this simplified model with
only one diagonal overall K\"ahler field $T$, the K\"ahler potential $K$ 
for 
any such field is given by $K=1/(T+T^*)$. The trilinear scalar soft terms 
are (see e.g. \cite{bimrev}) 
\beqa
A_{ijk}\ & =&\ h_{ijk}\left( F^SK_S\, +\, F^TK_T\, -\partial_T \log(K_i K_j K_k)
\right) \ 
=\ h_{ijk} F^SK_S\ = \ \\ \nonumber\  &=& -h_{ijk}
 {1 \over {M_p^2}}\, (S+S^*)^{-1/2}\, (T+T^*)^{-3/2}\ 
(\kappa_{10}^{-2})\int G_3^*\wedge \Omega \ =\ -h_{ijk}M^a \ . \nonumber
\label{tril}
\eeqa
Note that the computation of gaugino masses and trilinear terms reproduce 
our previous results obtained from the DBI and CS pieces of the D3-brane 
action. Indeed, in the limit or large volume $V$, the quantity $\kappa _{10} ^{-2}\int 
G_3^*\wedge \Omega$ becomes $\kappa _{10} ^{-2}G_0\times V =M_p^2 G_0$. 
On the other hand the prefactor scales like $g_s^{1/2}/(\sqrt{2}M_p^2)$,
 so that indeed one  obtains 
$|M|\rightarrow \frac {g_s^{1/2}}{\sqrt{2}}G_0$, in agreement
with the results obtained in section 2.

Hence we recover the fact that the $(3,0)$ component of $G_3$ corresponds 
to a vev for the auxiliary field of the dilaton, and reproduces soft terms 
of the dilaton domination kind; exactly as found in the explicit 
discussion from the D3-brane worldvolume viewpoint.

On the other hand, for  the scalar masses one obtains
\beqa
m_j{}^2\ &=&\ m_{3/2}^2\ +\ {{V_0}\over {M_p^2}}\ - \sum_i\, F^{\bar{i}}F^i\,
\partial _{\bar{i}}\partial _i \log(K_j) \ = \ \\ \nonumber
&=& m_{3/2}^2\ +\ {{V_0}\over {M_p^2}} \ -\ m_{3/2}^2\ =\
 { {\kappa_{10}^{-4}|\int G_{(3)}^*\wedge \Omega|^2} \over {
M_p^4\,  (S+S^*)\, (T+T^*)^3} } \, =\, |M_a|^2
\label{mascalar}
\eeqa
Notice that this does not agree with the DBI+CS result (\ref{dildom2}).
Indeed this is expected: Our local configuration in section \ref{dbics} 
assumed Poincare invariance in the four longitudinal directions. Hence 
the soft terms (\ref{dildom2}) correspond to those on configuration with 
$(3,0)$ flux and Poincare invariance. This cannot be reproduced in a flux 
compactification of the kind described by the above effective lagrangian, 
since non-ISD fluxes (like a $(3,0)$ component) always generate a non-zero 
vacuum energy (cosmological constant) that is not consistent with 4d 
Poincare invariance. In this sense, the result (7.11) should 
correspond to the soft terms obtained from the DBI+CS action for 
D3-branes in a modified background, allowing for deviations from Poincare 
invariance.

Notice that in heterotic analysis
\cite{il,kl,bim},  dilaton dominated SUSY-breaking by 
$F_S$ was argued to correspond to the soft terms (\ref{dildom2}). This 
however does not contradict our above comment, since in those heterotic 
references the vacuum energy was not computable in the underlying theory 
(since no microscopic SUSY-breaking source was proposed), and it was 
imposed by hand that it vanishes. If this were the situation, the ansatz 
in section \ref{dbics} would be a good local description, and soft terms 
would agree. However, in flux compactification the vacuum energy is 
computable and non-vanishing for non-ISD fluxes, leading to the above 
mentioned result.

Note that, if we insist in imposing the equations of motion in the flux 
compactification, one has $G_{(3)}^*\wedge \Omega=0$, which 
forces the ISD  condition, 
and we recover the well known fact that in no-scale models 
all soft terms identically vanish, to leading order \footnote{See 
footnote \ref{alphaprime}.}. One can also 
check that the field $T$ remains massless, whereas the complex dilaton 
field gets a mass equal to the gravitino mass (\ref{gravit}).

It is also important to emphasize that the analysis in section \ref{dbics} 
is more general than pure flux compactification of the kind in \cite{gkp}. 
For instance, one expects that a fully realistic model (involving 
additional ingredients) should reproduce a tiny cosmological constant 
(hence essentially enjoying 4d Poincare invariance) while the 
supersymmetry breaking sources (e.g. 3-form fluxes) are still large, and 
induce large soft masses. In this kind of model, the local description 
should be precisely of the form in section \ref{dbics}. This perfectly 
illustrates the usefulness of the analysis, and shows that 
the underlying difficulty to reconcile the local ansatz with a full-fledge 
compactification is the familiar difficulty to construct an explicit 
compactification with large SUSY-breaking and tiny vacuum energy.

\medskip

One could similarly analyze the 
SUSY-breaking due to vevs for auxiliary fields of other moduli, namely the 
complex structure moduli $M_a$. These are on an equal footing 
with respect to the dilaton in many aspects, but their couplings to the 
gauge sector fields is more model dependent, hence our discussion is more 
sketchy.

The complex structure moduli $M_a$ have a K\"ahler potential
\beqa
K(M,\bar{M}) & = & -\log \, (\, -i \, \int_X \Omega \wedge {\ov \Omega}\,)
\label{aversi}
\eeqa
with the dependence on the complex structure moduli is implicit in 
$\Omega$. Namely, for the complex structure modulus $M_a$, we have
$\partial_a \Omega  =  k_a \Omega\, +\, \chi_a$, where $k_a$ is moduli 
dependent, and the set of $\chi_a$ is a basis of $(2,1)$ forms. We then 
have
\beqa
\partial_a K   =  -k_a \quad \quad ;\quad \quad
\partial_a\partial_{\bar{b}}K  =  -{{\int_X \chi_a \wedge {\ov \chi_b}}
\over {\int_X \Omega \wedge {\ov \Omega}}}
\eeqa
Using the superpotential $W=\int_X G_3 \wedge \Omega$, the auxiliary field 
in the multiplet of $M_a$ is essentially given by 
\beqa
D_a W \, = \, \partial_a W+W\partial_a K \, =\, \int_X G_3\wedge \chi_a
\eeqa
Choosing the set of moduli to have diagonal metric, the auxiliary fields 
are very simple. The auxiliary field for $M_a$ is the piece of $G_3$ that 
wedges non-trivially with the corresponding $(2,1)$ form $\chi_a$. 
Clearly, only $(1,2)$ pieces of $G_3$ contribute to auxiliary fields of 
the complex structure moduli. In a general CY case the $(1,2)$ piece is 
primitive, and hence we see that only the IASD piece of $G_3$ contribute 
to the auxiliary field vevs. In $\IT^6$ or K3$\times \IT^2$, the 
non-primitive 
$(1,2)$ piece of $G_3$ seems to lead to vevs for auxiliary fields as well. 
Since these pieces are ISD, their contribution to the supergravity scalar 
potential must however be cancelled by additional negative pieces e.g. 
from the supergravity multiplet auxiliary fields, in analogy with the 
cancellation for $(0,3)$ components (indeed both cancellations are related 
by the enhanced supersymmetries of the underlying compactification), 
thereby preserving the no-scale structure (see \cite{ferrara,kors} for a 
discussion in $\IT^6$).

The computation of the explicit soft terms from this approach is again 
affected by the issue of the non-zero vacuum energy in compactifications 
with this kind of flux background. Hence it differs from the results from 
the local analysis in section \ref{iasdfluxes} by reasons already 
explained.

In conclusion, the low energy 4d supergravity approach allows us to 
identify the breaking of supersymmetry as a spontaneous breaking due to 
vevs for auxiliary fields of certain 4d supermultiplets.

\subsection{Soft terms and S-duality} 

The underlying Type IIB supergravity theory enjoys the  
property of $SL(2,{\bf Z})$ S-duality. Under transformations given in 
terms of a matrix $m=\pmatrix{a & b\cr c & d}$ with integer entries and 
unit determinant, the 3-form $G_{(3)}$ transforms as
\beq
G_{(3)}\ \rightarrow \ G_{(3)}'\ =\ { { G_{(3)}} \over {icS+d} }
\eeq
and the dilaton transforms as
\beq
iS\ \rightarrow \ { {iaS+b} \over {icS+d} }  \ .
\eeq
Interestingly the flux-induced superpotential transforms like a modular 
form of weight $-1$, i.e.
\beq
W(S)\ \rightarrow \ { {W(S)}\over {icS+d}}
\eeq
From this it is clear that the complete K\"ahler potential 
$K+\log|W(S)|^2$
is invariant under S-duality, since the transformation of the
$-\log(S+S^*)$ piece is cancelled by the transformation of 
the $S$-dependent superpotential. This works as in the first proposal
of the existence of S-duality in string theory in \cite{filq}.

Note in particular that due to this fact, all SUSY-breaking quantities, 
like the gravitino mass, are invariant under $D=4$ S-duality 
transformations, even though no unbroken SUSY in general remains. 
Similarly the SUSY-breaking soft terms felt by anti-D3-branes in the  
presence of $(0,3)$ backgrounds will be invariant as well.

\subsection{Finiteness properties of dilaton-domination
and holography}
\label{finiteness}

It was found by Jones, Mezincescu and Yao \cite{finite1} \footnote{For more 
recent discussion on these finiteness properties see e.g. 
\cite{finite2,finite3,finite4}.}, long time before 
the idea of dilaton domination was 
considered, that the choice of soft terms $A_{ijk}=-h_{ijk}M$ and $3m^2=M^2$ can be 
added to the $\NN=4$ finite theory without spoiling finiteness.
 In \cite{im,finite2,strings95} it was noticed that dilaton domination 
soft terms precisely corresponded to those boundary conditions. Since 
then it has remained unclear why dilaton domination soft terms have this 
property.

Our explicit realization of soft terms in string theory interestingly 
provides a new insight into this question. Consider a configuration of 
purely $(3,0)$ $G_3$ flux background in flat space. The full  supergravity 
solution (exact when regarded as an expansion in the flux density) is, 
as discussed in section (\ref{dbics}), of the black 3-brane form 
(3.9).

Notice in particular that the IIB complex coupling does not vary over 
spacetime. Now introduce a stack of D3-branes probing the  background. 
From the point of 
view of the 4d worldvolume gauge field theory, the flux background 
induces supersymmetry breaking soft terms of the dilaton-dominated kind 
(i.e. supersymmetry is broken by a vev for the auxiliary field which is 
partner of the dilaton). As we said, these soft terms 
preserve the finiteness of the theory. Moreover, the above results suggest 
that the gauge coupling is a marginal operator, on which the type IIB 
$SL(2,\IZ)$ acts as the S-duality group, leaving soft terms invariant.

The string theory construction allows to provide a simple interpretation 
for this result. It is known that the dependence of the IIB complex 
coupling on the coordinates transverse to the D3-brane 
reproduces the 
running of the field theory gauge coupling constant with the 
renormalization group scale. Hence in our situation 
the constant value of the dilaton suggests the softly broken theory still 
is finite, at least has vanishing beta function in perturbation theory. 
Moreover, the string theory construction captures the marginality of the
coupling parameter and the action of S-duality.

A more precise form of the statement could be achieved by using the 
AdS/CFT description of the system, namely taking the near horizon limit 
of the configuration of D3-branes in the flux background. However the full 
supergravity solution for the combined system is not known so it is 
difficult to extract precise conclusions at this point.

\medskip

We would like to finally point out an interesting observation. It was 
shown in \cite{finite3,finite4} that the finiteness condition for the softly 
broken $\NN=4$ theory are slightly more general, and read
\beqa
m_1^2+m_2^2+m_3^2 = M^2 \quad ; \quad A_{ijk}=-h_{ijk}M
\label{softfinite2}
\eeqa
where the $m_i$ are soft masses for the complex scalars in the three $\NN=1$ 
chiral multiplets in $\NN=4$. Hence, the finiteness property only constrains the 
trace of the mass matrix and allows the individual eigenvalues to change.

From our discussion after (\ref{genmass2}) and in appendix \ref{massesps}, it 
is clear that this kind of deviation from mass universality (without 
change in the trace of the mass matrix) is present if the configuration 
contains additional sources, located away from the D3-branes, besides the 
fluxes themselves. For instance, soft terms of the form 
(\ref{softfinite2}) arise in configurations with pure $(3,0)$ $G_3$ flux, 
and a set of anti-D3-branes located away form the D3-branes. The warp and 
5-form background created by the $\ov{\rm D3}$-branes superimposes over 
the 
flux background, and leads to changes in the scalar masses, keeping the 
trace of the mass matrix fixed (to the value $M^2$). It would be 
interesting to describe the finiteness of the resulting theory using 
holographic ideas in this more general backgrounds.

\section{A realistic orientifold with flux-induced soft terms}
\label{realistic}

As we have emphasized, the method presented to compute soft terms may be 
applied in a variety of situations. In particular, one can consider 
explicit compactifications where the 
above structures of soft terms are realized. A large class of type IIB 
Calabi-Yau or F-theory compactifications was introduced in \cite{gkp}, 
with O3-planes, D3-branes and ISD fluxes. From the viewpoint of soft 
terms this is a bit unfortunate, since as mentioned above ISD fluxes lead 
to cancellations between the DBI and CS parts of the D3-brane action, and 
produce no soft terms whatsoever. This agrees with the no-scale structure 
of the models.

One can still use these compactifications to study soft terms induced on 
the worldvolume of anti D3-branes. As discussed above, ISD fluxes do 
generate non-trivial soft terms on the latter. In fact, if the flux is 
purely $(0,3)$ the soft terms on the $\ov{\rm D3}$-brane are of the 
dilaton-domination type. These $\ov{\rm D3}$-branes can be regarded as 
probes of the configuration \footnote{One may worry about the 4-form 
tadpole they generate; one can assume we simultaneously introduce an 
additional D3-brane somewhere else in the internal space. Alternatively, 
one may introduce just $\ov{\rm D3}$-branes, and fluxes, see the main 
text.}. Interestingly this kind of situation also arises in the recent 
constructions of de Sitter vacua in string theory  \cite{kklt} 
(see also \cite{quevedo} for further developments), where the 
compactification includes fluxes that overshoot the RR tadpole from the 
background (e.g. O3-planes), and is compensated by the introduction of 
$\ov{\rm D3}$-branes (which are also responsible for lifting the vacuum 
energy and leading to the de Sitter vacuum).

In the following we describe an explicit example, in which we have a 
local configuration with a stack of anti-D3-branes on top of a $\IZ_3$ 
singularity. The resulting theory has a realistic Left-Right symmetric 
$SU(3)\times SU(2)_L\times SU(2)_R\times U(1)_{B-L}$ gauge group on the 
worldvolume of the antibranes
\footnote{It is equally easy to obtain similar models 
with the SM gauge group, see ref.\cite{aiqu}.
 We have chosen here a left-right symmetric
model  because the Chan-Paton matrices are simpler in this case.}. In 
order to cancel 
RR twisted tadpoles at this orbifold point
we introduce an additional stack of anti-D7-branes passing through the 
singularity. The resulting theory has three quark-lepton generations and 
three sets of Higgs doublets \footnote{The model is a variant of that 
presented in section 4.2.1 of the second ref.\cite{aiqu}.  In addition ISD 
fluxes are introduced as in appendix A in \cite{cascur}.}. Moreover soft 
terms of the dilaton-dominated type appear.

The model is a simple modification of the left-right symmetric
model in section 4.2.1 of the second ref.\cite{aiqu}.  
One just replaces branes by antibranes in that model,
except for the $D3$-branes at the origin which remain unaltered.
Untwisted tadpole cancellation will require the presence of
3-form fluxes.
We briefly review its construction in what follows.

Consider type IIB compactified on $\IT^6/\IZ_3$, where the orbifold is 
generated by the twist
\beqa
\theta\quad : \quad (z_1,z_2,z_3) \longrightarrow (e^{2\pi i/3} z_1, 
e^{2\pi i/3} z_2, e^{-4\pi i/3} z_3)
\eeqa
Let us mod out by the orientifold action $\Omega R (-1)^{F_L}$, where 
$R:z_i\to -z_i$, and introduce a $G_3$ flux of the form 
\beqa
G_3 & = & A\, d{\ov z}_1 d{\ov z}_2 d{\ov z}_3
\eeqa
which fixes the dilaton vev to $\tau=iS=e^{2\pi i/3}$. The coefficient $A$ 
is 
an even number to ensure proper quantization over 
toroidal 3-cycles \footnote{The space $\IT^6/\IZ_3$ does not have 
collapsed 3-cycles, hence there are no further subtleties in the 
flux quantization conditions, as compared with the toroidal case.}.
The flux is purely $(0,3)$ and breaks supersymmetry, but is ISD and obeys 
the equations of motion. Its contribution to the 4-form tadpole is 
$N_{\rm flux}=3|A|^2$. 

\begin{figure}
\begin{center}
\centering
\epsfysize=10cm
\leavevmode
\epsfbox{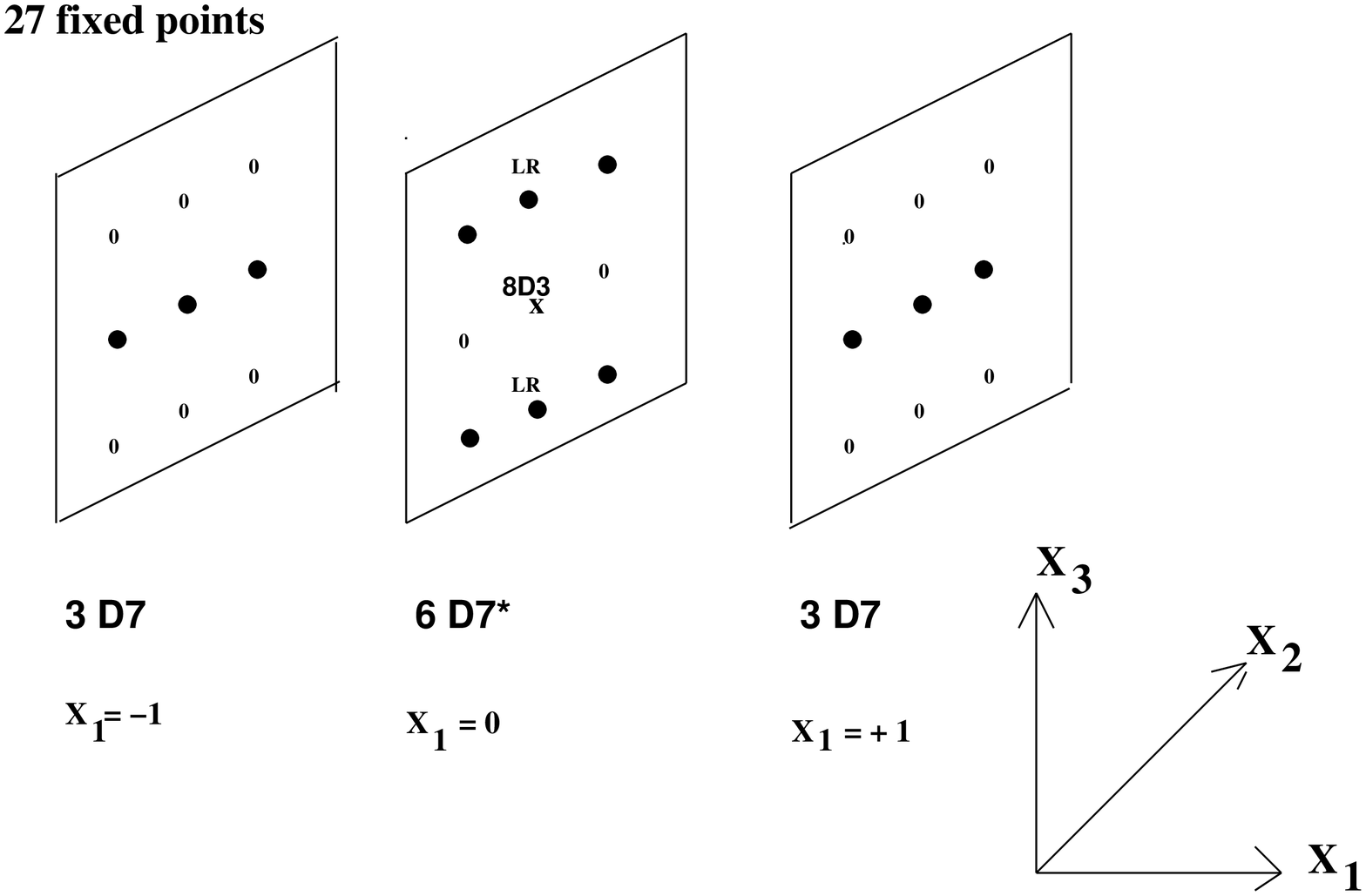}
\end{center}
\caption[]{\small A compact Type IIB $T^6/\IZ_3$ orientifold  model with a
three generation $SU(3)\times SU(2)_L\times SU(2)_R\times U(1)_{B-L}$
gauge theory. Here $X_i$ represent the three complex compact dimensions. 
The gauge theory lives on the worldvolume of
7 $\ov{\rm D3}$'s  located at the fixed point marked LR
(and its orientifold mirror). The rest of the  black dots represent one
$\ov{\rm D3}$.   
Upon switching on a self-dual
$(0,3)$ flux RR-tadpoles cancel and 
SUSY-breaking soft terms appear on the worldvolume of
the LR branes.
}
\label{fig1}
\end{figure}

There are 27 fixed points, which we label by  $(m,n,p)$, where 
$m,n,p=0,\pm 1$, as shorthand for the three possible positions of the 
fixed points in each complex plane. At the fixed point 
$(0,1,0)$ (and its orientifold mirror $(0,-1,0)$) we locate
7 $\ov{\rm D3}$'s  with Chan-Paton (CP) matrices:
\beq
\gamma_{\theta,{\bar 3}}\ = \diag \left( \id_3, \alpha\  \id_2, \alpha^2\
\id_2\right)
\label{cporientfour}
\eeq
(see fig.\ref{fig1}). In order to cancel twisted tadpoles at the fixed 
points $(0,n,p)$ we add 6 anti-D7-branes passing through them, with CP 
matrix
\beq
\gamma_{\theta,{\bar 7}}\ =\ \diag\left(\id_2, \alpha\ \id_2, \alpha^2\ 
\id_2\right)
\label{cporienttwo}
\eeq
We furthermore add a Wilson line $\gamma_W$ on them on the second complex 
plane, given by 
\beq
\gamma_{W,{\bar 7}}\ = \ \diag\left(\alpha, \alpha^2, \id_2, \id_2\right)
\label{cporientthree}
\eeq
In this way the gauge group coming from the $\ov{\rm D7}$'s is broken to
$U(2)\times U(1)$ and the different fixed points have now different 
$\ov{\rm D7}$-brane CP matrices. We complete a consistent configuration, 
cancelling all RR twisted tadpoles, as follows (see fig.\ref{fig1}). We 
locate one $\ov{\rm D3}$-brane at each of the four fixed points  $(0, \pm 
1,\pm 
1)$ with  CP matrix $\gamma_{\theta,\bar{3}}=1$, and 8 D3-branes at the 
origin, with CP matrix $\gamma_{\theta,3} =  \diag \left(\alpha \id_4, 
\alpha^2 \id_4\right)$. Finally, in order to cancel global RR tadpoles, we 
add 3 $\ov{\rm D3}$'s as well as 3 parallel $D7$-branes (and their 
orientifold mirrors)  passing respectively 
through the fixed points of type $(1,n,p)$
and $(-1,n,p)$ (see ref.\cite{aiqu} for details).
One can easily check that all twisted tadpoles cancel in this configuration.

The total untwisted RR 4-form charge in the configuration is
\beq
Q_{RR}\ =\ -32(O3)-24(D3^*)+8(D3)\ =\ -48
\eeq
This charge is neatly cancelled if we add ISD $(0,3)$ flux with $A=4$, which 
contributes to the RR-charge $3\times (4)^2= 48$ units.

This brane configuration is (meta)stable. 
The $D7$ and $\ov{\rm D7}$'s are stabilized on the 
planes passing through the orbifold points. They are forced
to remain there in order to maintain twisted RR tadpole
cancellation. Equivalently, the orbifold projection removes the scalar 
associated to flat directions describing brane motion. 
This will avoid $D7$- $\ov{\rm D7}$ annihilation. The addition of fluxes 
will not destabilize them, since their only effect would be to generate a 
potential for those scalars, if they were present.

The same happens for the $\ov{\rm D3}$'s whose scalars get masses of the 
dilaton dominated type. The D3-branes at the origin do not get masses 
from fluxes, since they feel the ISD flux which gives no soft terms (to 
leading order) for them. They are however stuck at the origin again by the 
twisted RR tadpole conditions, which would be violated if any of those 
branes traveled to the bulk (equivalently, their worldvolume field theory 
does not contain scalars parametrizing the possibility of moving away the 
D3-branes). In fact the four ${\overline {D3}}$'s at the four $(0, \pm 
1,\pm 1)$ fixed points are also stuck by twisted tadpole conditions.
All in all, the whole brane configuration is (meta)stable due
to a combination of trapping at the fixed points and flux-induced scalar 
potentials.

The complete spectrum of the left-right symmetric model of interest is 
given in section  4.2.1 of ref.\cite{aiqu}. The gauge group is 
$U(3)\times U(2)\times U(2)$, with two anomalous $U(1)$'s being actually 
massive, and the diagonal combination giving $B-L$. In the 
${\bf {\overline {33}}}$ sector, we obtain matter fields
\beqa
{\bf {\overline {33}}}\,\, {\rm sector:} \quad \quad
3\ \left[ (3,\bar 2 , 1)
+ (\bar 3 , 1, 2)  +  (1, 2, \bar 2 )\right]
\eeqa
corresponding to three left(right)-handed quarks $Q_L^a(Q_R^a)$, $a=1,2,3$,
and three sets of standard Higgs multiplets $H^a$.  
From the  ${\bf {\overline {37}}}$  and  ${\bf {\overline {73}}}$  sectors, 
one gets:
\beqa
{\bf {\overline {37}}}\,\, {\rm sector:}\quad\quad
(3, 1, 1;1)_{-1}+(3,1,1; 2)_0+
(1,2,1;1)_{1}+ (1,2,1;2)_0 \nonumber \\
{\bf {\overline {73}}}\,\,{\rm sector:}\quad\quad 
(\bar 3, 1,1; 1)_1 + (\bar 3,1,1; 2)_0 + (1,1,2;1)_1+(1,1,2;2)_0
\eeqa
These contains three left (right)-handed leptons $L^a(R^a)$. There are also  
some extra vector-like pairs of colour triplets will in general become 
massive once $(7_i7_i)$ states get vevs (see ref.\cite{aiqu}). The 
orientifold projection map the sets of branes at $(0, 1, 0 )$ and 
$(0, -1, 0)$ fixed points to each other, so only one copy of the LR model 
is obtained.

The quarks in this model have a superpotential
\beq
W_{Y}\ =\ g\sqrt{2} \epsilon_{abc}Q_L^aQ_R^bH^c
\label{yuk}
\eeq
On the other hand there are no renormalizable lepton Yukawas
which may only appear after a blowing up of the singularity \cite{aiqu}.
We will thus concentrate here on the quarks.
The flux background is of ISD $(0,3)$ type, hence it leads to dilaton 
dominated soft terms on the worldvolume of the anti-D3-branes. As 
discussed in section \ref{iasdfluxes}, these are 
\beqa
 m_{Q_L^a}^2 \ =\  m_{Q_R^a}^2 \ =\  m_{H^a}^2 \ &=&\ m_a^2 \ \ a=1,2,3 
\nonumber \\
 m_{1}^2 + m_2^2 + m_3^2 \ &=&\
\frac{g_s}{6}\vert G_{{\bar 1}{\bar 2}{\bar 3}} \vert^2 \nonumber \\
M_3=M_L=M_R=M_{B-L}=M\ &=&  \  {{g_s^{1/2}}\over {\sqrt{2}}}\
G_{{\bar 1}{\bar 2}{\bar 3}} \nonumber \\
A^{abc}\ &=& \ - h^{abc}M
\eeqa
with the rest of the soft terms vanishing. This kind of 
SUSY-breaking soft terms applied to the MSSM have been abundantly studied 
in the literature, and provides a phenomenologically interesting and 
viable soft term pattern \cite{blm,bim,bimu,clm,aaqik}. 

In the above analysis  we have ignored  the effect of the closed string 
K\"ahler moduli, which are not stabilized in the present context, and 
would acquire runaway potentials due to the uncancelled tensions. This 
feature would clearly not be present in more complete models which could 
include a mechanism to stabilize K\"ahler moduli, like gauge 
non-perturbative effects \cite{kklt},
and a small cosmological constant. 
In that case configurations of the 
kind presented above would be metastable. Still, the soft terms felt by the SM 
fields would be essentially independent of the avatars of the 
stabilization, and would be  essentially of the dilaton domination form 
shown above.

\medskip

Clearly the above kind of construction can be carried out in other 
manifolds, and configurations. It is interesting that the general pattern 
applies to models closely related to the class of models leading to 
de Sitter compactifications.

\section{Some phenomenological issues}
\label{pheno}

\subsection{Mass scales and the hierarchy}

The flux quantization conditions  tell  us that on any 3-cycle $\Pi$
\beq
\frac{1}{2\pi\alpha'}\int _{\Pi} \ F_3 \in 2\pi{\IZ} \quad ;\quad
\frac{1}{2\pi\alpha'}\int _{\Pi} \ H_3 \in 2\pi{\IZ}
\eeq
As discussed in \cite{gkp}, compactifications with 3-form
fluxes do not stabilize the overall volume modulus $T$. Therefore the 
very
low energy effective field theory generically contains it as a modulus, 
and its parameters are functions of it. In particular, soft terms depend 
on the volume modulus because, due to the above quantization conditions, 
the  flux densities have the dependence for large radius
\beq
G_3 \ = \ f{{\alpha '}\over {R^3}}
\eeq
with $f$ an $R$-independent constant measuring the amount of quantized 
flux.  This is  the well know fact that flux effects are suppressed like
$1/R^3$. From previous sections we see that the typical scale
for soft terms will then be
\beq
m_{soft}\ =\ {{g_s^{1/2}}\over \sqrt{2}}G_3\ =\ {{fg_s^{1/2}}\over \sqrt{2}}
{{\alpha '}\over {R^3}}
\label{msoft}
\eeq
This could be interesting because for large radii   
we can get a hierarchy of scales 
\beq
 {g_s^{1/2}}{{\alpha '}\over {R^3}} ({\rm fluxes})\  \ll 
{1\over R}({\rm KK}) \ \ll \  M_{s}\ 
\eeq
(with $M_s$ the string scale) and identify the lightest one with the
electroweak scale. For sufficiently large R the tower of KK states may be
ignored and we can discuss physics of fluxes using the 4d effective
field theory. This has been implicit in our discussion in section 
\ref{sugra}. On the other hand, the dependence of the Planck scale with 
the overall volume is given by
\beq
M_p \ =\ (\alpha ')^{-2} R^3
\eeq
and hence we have 
\beq
m_{soft} \ =\ {{f\ M_s^2}\over {M_p}}
\label{geometrical}
\eeq
with $M_s={\alpha '}^{-1/2}$ the string scale.
This expression, in the effective $\NN=1$ supergravity 
language is consistent with some auxiliary field
(like e.g. that of the dilaton ) getting a vacuum expectation
value of order $fM_s^2$.  
Eq. (\ref{geometrical})  is an interesting expression. Note that if one wants 
to identify $m_{soft}$ with the electroweak scale, the string scale should 
be of order the intermediate scale, $M_s=10^{10}$ GeV, a possibility 
considered by some authors in the past \cite{intermediate}. Note that in 
order to get the correct value for $M_p$ the compact radii should be of 
order $R=10^3 \sqrt{\alpha'}$.

Note that in this scenario the SUSY-breaking scale is small compared
to the string scale because the flux is diluted by the fact the the
compactification radius is 3 orders of magnitude larger than the string
length scale. This  requires a choice
of large overall K\"ahler modulus vev. Notice however that there are 
interestingly models where the latter is stabilized via non-perturbative
effects, and used in the construction of de Sitter vacua in string theory
\cite{kklt} (see also \cite{quevedo}).  Stabilization at such moderately  
large overall volumes is achieved with small fine-tuning. It would be   
interesting to find explicit vacua in which these features arise.

Note also that the above computation involved only the dependence on the 
overall scale, and assumed more or less homogeneously diluted 3-form
fluxes. In particular it would be possible to consider situations where
the D3-branes are located in a region of the internal space where the
fluxes are hierarchically diluted as compared with other regions. One
prototypical example would be Calabi-Yau's containing throats in which 
most of the flux is concentrated
\cite{ks,gkp}. D3-branes away from the throat would 
feel the flux in a very suppressed manner, and the soft terms would be 
affected by additional suppression factors, as compared with the above 
estimate. This would correspond to a suppressed coefficient $f$ in 
eq.(\ref{geometrical}). In this case part (or  all)  of the
weak scale versus Planck scale suppression could be due to this effect 
and the string scale could be closer (or identified) 
with the Planck scale.
On the other hand the inverse situation is possible. If D-branes are 
located in a region of compact space with much more flux than the average,
the string scale could be lower than the intermediate scale. In this 
setup, the largeness of the Planck mass would not be due to a large size 
for the extra dimensions, but rather to their strong warping 
\cite{rs,gkp}. 

Let us finally comment that in the early days of heterotic, it was soon 
realized that a 3-form flux background for $H_{ijk}$ would break 
supersymmetry \cite{din,drsw}. 
However it was noted in \cite{rohmwitten} that its flux was quantized 
\footnote{See \cite{kachruhet} for a recent improvement of the situation.}
so that the scale of SUSY-breaking would be necessarily of order of the 
Planck scale (recall that in the perturbative heterotic case $M_p$ is 
directly related to $M_S$ and that making compact radii very large leads 
to a difficult regime of strong coupling). Type II compactifications 
provide the new possibility discussed above, compact radii may be large 
in a consistent manner, and D3-branes lead to localized gauge sectors, 
sensitive to the flux density rather than to the integral of the flux. 
Thus the setup allows
for hierarchically small breaking of SUSY, if desired. 

\subsection{The structure of soft terms and its application to the
SUSY SM}

It is conceivable that the soft terms of a completely
realistic version of the MSSM embedded in a Type IIB vacuum 
could have their origin in the supergravity backgrounds
that we are considering.  The details of those terms would be 
model dependent, but one can advance some general
properties of the soft terms that one would expect to appear
if SUSY-breaking has the origin considered in this paper.
Some general properties are the following: \\

{\it  i) Gaugino masses and trilinear $A$-terms}

These  only appear (in leading order) if a background for $G_3$ is 
present. This is important for model-building since present limits
on SUSY particles already require the presence of substantial
gaugino masses. The relationship $A=-hM$ found from flux induced
SUSY-breaking is quite model independent. In as much as one can associate 
the heaviest SM generation Yukawa coupling as coming from a truncation of 
an underlying $\NN=4$ theory (which happens in many realistic 
compactifications constructed up to now) one would expect the constraint  
$A_t=-h_tM$ to be quite generic. Here $A_t$ is the trilinear term 
associated to the top quark Yukawa coupling. \\

{\it ii) Scalar $m^2$ terms}

 Whereas gaugino masses and trilinear terms  depend only on the
local background $G_3$ around the location of the branes,
scalar masses also depend on other backgrounds (5-form and dilaton)
as well as on the form of the warping. Hence they are 
sensitive to other distant sources (like other branes or
antibranes, orientifold planes etc.). Thus scalar masses
will depend on details of the full vacuum solution.

In general,  soft scalar masses are different for different scalar fields. 
Thus in a realistic compactifications different e.g., squark flavors will 
have different masses. However, for configurations and backgrounds with 
certain  geometrical symmetries (like the spherically symmetric cases 
mentioned e.g. in subsection \ref{iasdfluxes}, leading to dilaton 
dominated type of soft terms) scalars may all have identical masses.
This issue is important phenomenologically, since there are strong 
constraints on mass differences, e.g. among the first and second 
generation squarks in the MSSM, coming from sufficient suppression of 
Flavor 
Changing Neutral Currents (FCNC).

Note that distant sources like e.g., antibranes give rise on the 
worldvolume of branes to D-term type of breaking, i.e. scalar masses but 
neither gaugino masses nor trilinear terms. This is in general  
insufficient for getting a phenomenologically acceptable SUSY-breaking. \\

{\it iii) $\mu $-term}

As we saw in the specific $\IZ_4$ example in section \ref{orbifolds}, 
fluxes provide a new way to obtain a `$\mu $-term' in string theory. The 
obtained term is proportional to the primitive component of $(1,2)$ 
antisymmetric fluxes ($S_{ij}$ components). It is natural in flux induced 
SUSY-breaking that all fluxes (either e.g. $(3,0)$ leading to gaugino 
masses and $A$-parameters as well as $(1,2)$ type, leading to a $\mu 
$-term) to be of the same order $\propto g_s^{1/2}\alpha '/R^3$. Thus the 
present scheme provides a simple solution to the so called `$\mu$-problem', 
which is understanding why a SUSY-term like $\mu$ is of the same order of 
magnitude as SUSY-breaking soft terms. They simply have similar origin. \\

{\it iv) B-term}

The present scheme gives also a natural source for the 
presence of a B-term in the theory. It depends on the
warping, 5-form and dilaton backgrounds and hence it is
more model-dependent than the rest of the terms. Furthermore,
its value is not fixed by the value of antisymmetric 
fluxes via the local supergravity equations of motion.
Nevertheless one would expect its size to be of the 
order of the other soft terms, as can be estimated in explicit examples. 
\\

{\it v) Complex phases }

In general the soft $A$, $M_a$, $\mu$ and $B$ parameters are complex. The 
phases of the first three are related to the (generically  complex) $G_3$ 
backgrounds. In some particular cases (like e.g. when $G_3=G_3^*$, 
corresponding to vanishing NSNS 3-form flux) they may become real. The $B$
parameter is generically complex and unrelated to the flux phases. All 
these phases are relevant for the so called SUSY CP-problem, which is that 
arbitrary complex soft terms give rise to CP-violating transitions 
(electric dipole moment of the neutron), with rates three orders of 
magnitude larger  
than the experimental bounds. In the MSSM the relevant phases
for these transitions are $\phi_A={\rm arg}(M_aA^*)$  and 
$\phi_B={\rm arg}(MB^*)$. Note that in fluxed induced SUSY-breaking one has
$\phi _A=0$ and there is no CP-violating contribution from that
source. On the other hand the value of $\phi _B$ will depend
on the phases of $B$ and $M_a$ which are in general uncorrelated.
On the other hand, by appropriately choosing the
form of the warping, 5-form and dilaton backgrounds 
($K_{ij}, \chi _{ij}$  and $\pim\tau _{ij}$ in eq.(\ref{softgen}))
one can always get a vanishing $\phi _B$. \\

The above considerations seem to suggest the necessity of considering
explicit antisymmetric $G_3$ background and negligible effects from distance
sources on the SM particles, in order to get a phenomenologically
viable model. In order to obtain sufficient degeneracy among
squarks, some geometrical symmetry should be present.

\section{Final comments}
\label{conclu}

In the present paper we have computed the SUSY-breaking soft terms
on the worldvolume of D3-branes induced by a general class of
Type IIB supergravity backgrounds including RR and NS 3-form fluxes.
The supergravity backgrounds are expanded around the location of the
$D3$-branes, and the lowest terms in this expansion correspond 
precisely to  the lowest dimensional SUSY-breaking soft terms.
Different patterns of SUSY-breaking soft terms are obtained 
depending on the tensorial structure of the antisymmetric $G_3$
backgrounds. The results are summarized
in eqs.(\ref{softgen}). The computation applies to $D3$-branes sitting on
a smooth point in compact case but also to $D3$ branes sitting on
e.g., orbifold singularities leading to chiral theories with
phenomenological interest.

This approach is quite general in the sense that it allows  us to compute
SUSY-breaking terms induced by a general supergravity background. Thus 
for example, one can  compute the breaking induced by distant
${\ov {D3}}$-branes or orientifold planes.
It also applies to explicit compactifications in which one can treat
the physics around the $D3$-brane position locally. 
In compactifications the flux-induced SUSY-breaking terms may be
interpreted as arising from vevs of the auxiliary fields 
of the complex dilaton $S$, the K\"ahler and complex structure moduli.
This is somewhat  analogous  (though not identical) 
to the results for soft terms obtained in heterotic
\cite{il,kl,bim,bimrev} and Type I \cite{imr,kn}  schemes.
 While the soft terms obtained
from $(0,3)$  {\it or}  $(3,0)$  do indeed correspond to the
no-scale and dilaton domination soft terms in those references,
general combinations of ISD and IASD fluxes lead to new
possibilities not previously considered in heterotic schemes.
One of the new interesting aspects is the generation of both
$B$ and $\mu $-terms in a novel way.
 
One interesting question is the structure of mass scales 
introduced by flux SUSY-breaking.
In particular, one would like to identify the soft terms
with those required by the MSSM for its viability. We find that soft
masses are of order $M_s^2/M_p$, with $M_s$ the string scale and that 
in order to identify that with the 
 electroweak scale one  needs  to have
$M_s=10^{10}$ GeV. This is compatible with the observed 
value of the Planck  scale 
 if the compactification scale  is
three  orders of magnitude smaller than the string scale. 
Note that in this scheme two effects are the cause of the 
large hierarchy between electroweak and Planck scales.
A large compact volume increases the value of $M_p^2$
(which goes like $\kappa_{10}^{-2}\,{\rm Vol}$)
whereas at the same time dilutes the flux-induced 
soft terms (which go like $\alpha'/{\rm Vol}^{1/2}$).
The combination of both effects gives rise to a large 
hierarchy although, as we said, the compact radius 
is only 3 orders of magnitude larger than the string length.
On the other hand the presence of a large warp factor
could also lead to the generation of the hierarchy.

Another interesting property is the fact that all soft
terms so obtained are invariant under $D=4$,  $SL(2,{\bf Z})$
 S-duality acting on the complex dilaton and the 
three-form flux. One also gets an understanding of the 
finiteness of the dilaton-dominated soft terms 
when added to the $\NN=4$ lagrangian, which arises from
the fact that the dilaton is constant in the IASD 
$(3,0)$ background and holography.

Examples of CY compactifications with ISD $(0,3)$ fluxes 
 were  considered  in \cite{gkp},   
containing also  O3-planes and  D3-branes. 
Those compactifications have a no-scale structure which cancels 
soft terms to leading order. It is easy to modify those 
schemes in order to have non-vanishing (and interesting) 
SUSY-breaking soft terms by adding anti-D3-branes to the
compactification. The degrees of freedom on the worldvolume 
of ${\ov {D3}}$-branes feel effective $(3,0)$ fluxes, leading 
to non-trivial soft terms. We have constructed an explicit
$Z_3$ orientifold example with a $SU(3)\times SU(2)_L\times SU(2)_R\times
U(1)_{B-L}$ three generation model in which this effect takes place. 
We believe it should be possible to construct in this way
models in which all moduli are fixed yet non-trivial 
soft terms are obtained for the SM fields.

\medskip

\vspace*{0.5cm}

\medskip

{\bf \large Acknowledgments}

We thank J. F. G. Cascales, D. Cremades,  D.R.T. Jones
and  F. Quevedo for useful  discussions.
L.E.I. and P.G.C. thank CERN's Theory Division for hospitality.
This work has been partially supported by the European Commission under 
the  RTN contract HPRN-CT-2000-00148 and the CICYT (Spain).
The work of P.G.C. is supported by  the Ministerio de Educaci\'on, Cultura 
y Deporte (Spain) through a FPU grant.

\newpage

\appendix

\section{SUSY-breaking soft terms}

We include in this section for reference some general
definitions for soft terms which we employ in the main text.
We are interested in gauge theories which can be written 
in some approximation as $\NN=1$ SUSY theories with 
chiral fields $\phi _i$, $i=1,..M$ transforming under some
representation of the gauge group. Let the 
renormalizable superpotential 
have the general form
\beq
W(\phi)\ =\  
 {1\over 3!} h^{ijk}\phi_i\phi_j\phi_k
\eeq
The $\NN=4$ case is a particular example with 3 adjoint chiral fields.
There is a finite list of soft SUSY-breaking terms which may be added to 
the lagrangian without introducing  quadratic divergences. One has
\beq 
L_{soft}\  = \ -(m^2)_{ij}\phi_i \phi_j^* \ -\ 
( {1\over 3!} A^{ijk}\phi_i\phi_j\phi_k \ +\
{1\over 2} B^{ij}\phi_i\phi_j \ -\  {1\over 2} M^{a}\lambda_a\lambda_a \ 
+\ h.c.) \ . 
\label{soft1}
\eeq
Here $\lambda_a$ is a gaugino field and $a$ runs over the different gauge group 
factors.  One can also consider the addition of extra terms as in
\beq
L^{(2)}_{soft}\ = \ - {1\over 2}\mu ^{ij}\psi _i\psi _j \ +\
 {1\over 2} C^{ijk}\, \phi_i\phi_j^*\phi_k^* \ +\  M_g^{ia}\, 
\psi _i \lambda _a \ +\ h.c.
\label{soft2}    
\eeq 
where $\psi_j$ denotes the fermionic partners of the $\phi_j$.
The last term is only present if the matter field transforms in the 
adjoint. The other two terms may in fact lead to quadratic divergences
(and hence they would not be soft) but such quadratic divergences are
absent if the chiral multiplets are not singlet under the gauge group,
so in full generality one should consider them
\cite{jones}. In fact, in the
case
\beq
(m^2)_{ij}\ =\ |\mu ^{ij}|^2\ \ ;\ \ C^{ijk}\ =\ -(h^{jkl})^*\mu _{il}
\label{susyc}
\eeq
with the rest of the soft terms vanishing, the theory becomes supersymmetric
and is described by the above trilinear superpotential plus a 
a SUSY mass term ${1\over 2}\mu _{ij}\phi_i\phi_j$.  

There are some choices of $\NN=0$ soft terms which have particularly
interesting ultraviolet  properties. Consider in particular the case of
$\NN=4$ supersymmetry. It has been shown that the choice of soft terms
\beq
A^{ijk} \ =\ -M h^{ijk} \ \ ;\ \  Tr (m^2)_{ij} \ =\ |M|^2
\label{dilat}
\eeq
with the other terms vanishing is ultraviolet  finite to all
orders in perturbation theory
\cite{finite1,finite2,finite3,finite4}. It is interesting that if one assumes
that  the auxiliary field of the complex dilaton $S$ is the only
source of SUSY-breaking in a $D=4$ heterotic compactification
one gets the result:
\beq
A^{ijk} \ = \ -\sqrt{3}m_{3/2} h^{ijk} \ \ ;\ \ M_a\ =\ \sqrt{3}m_{3/2}
\ \   m_i^2\ =\  \ m^2_{3/2}
\label{tomaya}
\eeq
which is a particular case of eq.(\ref{dilat}). Thus dilaton dominated
SUSY-breaking applied to a $\NN=4$ theory maintains finiteness to all
orders. 

\medskip

\section{More general backgrounds}
\label{moregen}

\subsection{Linear terms}
\label{linear}

As discussed in the main text, our power-expansion of the background 
fields starts at the quadratic order. This is because linear terms in 
the metric, dilaton, $\chi_4$, ... generically lead to terms linear in the 
scalar fields in the D3-brane worldvolume action. In this appendix we 
briefly describe this in more detail.

For instance consider a background metric with a linear term, for instance
\beqa
Z_1^{1/2}(x^m)\, =\,  1\, +\,  Z_m\,  x^m\, +\ldots
\eeqa
From the DBI action, the leading terms in the worldvolume scalar action 
are
\beqa
{\cal L}\, =\, \frac{1}{2}\,\partial_\mu\phi^m\partial^\mu \phi^m\,
 +\, 2 \, Z_m\,  \phi^m
\eeqa
so that the configuration sits in the slope of the scalar potential. It 
may be possible to consider configurations where the linear terms in the 
background expansion cancel out at the level of the scalar potential, but 
this situation is highly non generic. 

It is easy to consider examples of backgrounds with such linear terms. 
Consider the D3-brane is embedded in the supergravity background created 
by a stack of anti D3-branes at a distant point $x_0^m$. The background is
\beqa
ds^2 & = & Z(x^m)^{-1/2}\, \eta_{\mu\nu} dx^\mu dx^\nu\, +\, Z(x^m)^{1/2} 
dx^mdx^m \nonumber\\
\chi_4 & = & \frac{1}{Z}\, dx^0 dx^1 dx^2 dx^3 \nonumber\\
F_5 & = & d\chi_4\, +\, *d\chi_4 \nonumber\\
Z & = & 1\,-\, \frac{1}{2\pi^2}\, \frac{N}{|\vec{x}-\vec{x_0}|^4}
\eeqa
We have 
\beqa
Z & = & {\rm const.}\,-\, \frac{2x_0^m N}{\pi^2 r_0^6} \, x^m \, +\,
\frac{N}{\pi^2 r_0^4} \, \left[\, 
\frac{\delta_{mn}}{r_0^2}\, +\, 6\frac{x_0^mx_0^n}{r_0^4}\,  \right]\, x^m 
x^n
\ldots
\eeqa
with $r_0^2=\sum_n (x_0^n)^2$. This leads to linear terms for $\phi^m$, making 
the configuration unstable. Recalling the interpretation of $\phi$ as 
coordinate of the D3-brane in transverse space, the instability 
makes the D3-brane move towards the point $\vec{x}_0$, i.e. towards the 
antibrane. Namely, this is the worldvolume derivation of the 
gravitational attraction felt by the D3-brane. There is a similar linear 
terms arising from the D3-brane interaction with the RR 5-form background 
created by the antibranes

\subsection{Scalar masses}
\label{massesps}

In this section we elaborate on the fact that the warp factor is not fully 
determined in terms of the fluxes unless some choice of boundary 
conditions is made. For instance, it is clear that to any solution for 
the equation of motion (\ref{ricci}), (\ref{tracemetric}) one can 
superimpose a solution of the homogeneous equations. For 
instance, we can superimpose the 
\beqa
Z & = & {\rm const.}\, -\, 
\frac{1}{2\pi^2}\, \frac{1}{|\vec{x}-\vec{x_0}|^4} \,
-\, \frac{1}{2\pi^2}\, \frac{1}{|\vec{x}+\vec{x_0}|^4}\, = \nonumber \\
& = & {\rm const.}\, +\, \frac{2}{\pi^2 r_0^4}\, \left[\, 
\frac{\delta_{mn}}{r_0^2}\, +\,6\, \frac{x_0^mx_0^n}{r_0^4}\, \right]\, 
x^m x^n
\eeqa
In the patch $|\vec{x}|<r_0$ the above is a vacuum solution. The 
additional piece in the warp factor can be regarded as the metric 
background created by a pair of antibranes located at $\pm \vec{x_0}$ (so 
that the linear terms cancel out due to the $\IZ_2$ symmetric 
distribution), but it can be regarded as a vacuum solution in the local 
patch around the D3-branes (at $\vec{x}=0$). In the latter view, the 
additional piece of the warp factor modifies the behaviour of the metric 
background at the boundary of the patch. 

In any of these views, it is clear that the additional background modifies 
the masses of the worldvolume scalars, by an amount which is completely 
unrelated to the 3-form fluxes present in the configuration. Notice that 
this is consistent with the equation of motion, and so implies that the 
additional piece of the mass matrix is traceless.

Hence, as mentioned in the main text, the expression of the scalar masses 
in terms of the background metric, dilaton and 5-form depends just on the 
local background for these fields. Now specifying a relation between the 
scalar masses and the fluxes involves a choice of boundary condition for
the metric, dilaton and 5-form. In this sense, locality is more manifest 
in the expression of soft scalar masses in terms of the metric, dilaton, 
and 5-forms, without use of the equations of motion.

\section{Conventions}
\label{conventions}

Throughout the paper we use a perturbed $SO(9,1)$ Minkowskian
metric with `mostly plus' signature, and antisymmetric tensor defined 
by
$\epsilon_{0123456789}=1$. Indices $M,N,...$ are 10d, indices
$\mu,\nu,...$ are 4d, and indices $i,j,...$ run over the six dimensions
transverse to the D3-brane.

In general we take the same conventions as in \cite{gkp,pol}.
Hence, the type IIB supergravity action is
given by (already in the Einstein frame)
\beqa
S_{IIB} & = & \frac{1}{2\kappa^2}\, \int d^{10}x\, (-G)^{1/2}\,
[\, R - \frac{e^{2\phi}}{2}\partial_M\tau \partial^M \bar{\tau}
-\frac{1}{12}e^{\phi}G_3\cdot \bar{G}_3-\frac{1}{480}F_5^2\, ]\, + 
\nonumber\\
&& + \, \frac{1}{8i\kappa^2}\,
\int\, e^{\phi} C_4 \wedge G_3 \wedge \bar{G}_3
\eeqa
where
\begin{eqnarray}
 G_3= F_3 - \tau H_3\ & & \quad F_3= d C_2 \\
F_5= dC_4 - \frac{1}{2}C_2 \wedge H_3 + \frac{1}{2}B_2\wedge F_3\ &
& \quad \tau = C+ie^{-\phi}\ .
\label{fcinco}
\end{eqnarray}
Note that $C_4$ defined by (\ref{fcinco}) in general is
different from $\chi_4$, defined in
(\ref{ansatz}). However, since in our particular background
$B_2$ and $C_2$ vanish in the
worldvolume, we have that $(C_4)_{0123}=(\chi_4)_{0123}$.

From here we can derive the equations of motion and Bianchi
identities. In particular, we will use the following ones
\begin{eqnarray}
\label{eomfirst}
dF_5 & = & \frac{i}{2} g_sG_3 \wedge \overline{G}_3 \\
\bigtriangledown^M \left(\frac{i}{2} e^{\phi} \partial_M \tau 
\right)&-&
\frac{1}{2}e^{2\phi} \partial^M C \partial_M \tau =
\frac{g_s}{24}G_{MNP}G^{MNP}\, \\
R_{MN}&=& \frac{1}{4} e^{2\phi} \partial_M\tau \partial_N\tau^\ast +
\frac{1}{4} e^{2\phi} \partial_M\tau^\ast \partial_N\tau +
\frac{g_s^2}{96}F_{MQRST}F_N~^{QRST}
\nonumber\\
&&+\frac{g_s}{8}\left({G_M}^{PQ}G_{NPQ}^\ast +{G_N}^{PQ}G_{MPQ}^\ast-
\frac{1}{6}g_{MN}G_{PQR}^\ast G^{PQR}\right)\\
d*\preal G_3 &=& F_5\wedge H_3 \, .
\label{eomcomplete}
\end{eqnarray}
where our prescription for the hodge dual operation * is the common one 
in supergravity
\begin{eqnarray}
*A_{\mu_1\ldots\mu_{d-p}}=\frac{1}{p!}\epsilon_{\mu_1\ldots\mu_{d-p}}   
{}^{\nu_1\ldots\nu_p}A_{\nu_1\ldots\nu_p}
\end{eqnarray}

Note that our conventions are slightly different from the ones of 
\cite{myers}.
In particular
\beqa
{\rm Myers} & & {\rm Ours} \nonumber \\
C_4 & \quad \to \quad & C_4+(1/2)B_2\wedge C_2 \\ \nonumber
B_2 & \quad \to \quad & -B_2
\eeqa
Hence Myers' action takes now the form
\begin{eqnarray}
S=& -\mu_3 \int d^4 \xi \, \Tr \left[ \, e^{-\phi}\,
\sqrt{-\det(P[E_{\mu\nu}+E_{\mu i}(Q^{-1}-\delta)^{ij}E_{j\nu}]\,
+\, 2\pi\alpha'\, F_{\mu\nu})\, \det(Q_{ij})} \, +\right] \nonumber\\
& +\, \mu_3\, \int \Tr \left(\, P\left[e^{2\pi i\,\alpha'
{\bf i}_{\phi}{\bf i}_{\phi}} \,
(\, \sum_n C_n+\frac{1}{2}B_2\wedge C_2\, ) \, e^{-B}\, \right]\, 
e^{2\pi\alpha'F}\,
\right)
\end{eqnarray}
with the usual definitions for the pull-back, $Q_i{}^j$ and
\begin{eqnarray}
E_{MN}=G_{MN}-B_{MN}
\end{eqnarray}
The new term $B_2\wedge C_2$
will   make no difference in our computations since its  background
 will be null in the worldvolume in order
not to violate Lorentz invariance.

We hope there is no confusion between the momentary double use of 
$\phi$, as in the dilaton $e^{-\phi}$, and as the worldvolume scalars
$\phi_m$ in ${\bf i}_{\phi}$. In the main text, $\phi$ is used only for
the latter purpose.

Consistently with the above, it is adequate to define $C_6$ through
\footnote{See \cite{ps} for further details.}
\begin{eqnarray}
-*\preal G^{(3)} = dC_6 - H_3\wedge \left(C_4+\frac{1}{2}B_2\wedge 
C_2\right)
\label{relationc6}
\end{eqnarray}
In fact, we see that when we apply exterior derivative to 
(\ref{relationc6}) we
recover the corresponding equation of motion (\ref{eomcomplete}).

Whenever we complexify the transverse 6 dimensions we will do it by 
means of
\begin{eqnarray}   
z^l = \frac{1}{\sqrt{2}}(x^{2l+2} + ix^{2l+3}) \\
\bar{z}^l = \frac{1}{\sqrt{2}}(x^{2l+2}-ix^{2l+3})
\end{eqnarray}

It is not hard to see that our metric, antisymmetric and delta tensors 
on
this new basis take the form
\begin{eqnarray}
g_{\mu\nu}&=&\diag(-1,1,1,1)=-\eta_{\mu\nu} \nonumber\\
\epsilon_{\tilde{0}\tilde{1}\tilde{2}\tilde{3}1\bar{1}2\bar{2}3\bar{3}}
&=& (-i)^3 = i \quad ; \quad
\epsilon_{1\bar{1}2\bar{2}3\bar{3}} = i
\quad ; \quad
\epsilon_{123} = \epsilon_{\bar{1}\bar{2}\bar{3}} = 1 \nonumber\\
g_{ij} &=& g_{\bar{i}\bar{\j}} = 0 \quad ; \quad
g_{i\bar{\j}} = g_{\bar{i}j} = 1 \nonumber\\
\delta_{i\bar{i}} &=& \delta_{\bar{i}i} = 1 \quad ; \quad
\delta_{ii} = \delta_{\bar{i}\bar{i}} = 0
\label{tenscmp}
\end{eqnarray}
where tilded indices mean worldvolume indices.

Note that $g_{\mu\nu}$ in (\ref{tenscmp}) does not correspond to the 
usual choice in field
theory where $\eta_{\mu\nu}=diag(1,-1,-1,-1)$. Therefore, in the low
energy effective action we must change the sign of the contractions of
the form $A_{\mu}A^{\mu}$.
 
The complex conjugate tensor is defined in complex coordinates by
\begin{equation}
A_i^*=(A_{\bar{i}})^* \quad ; \quad A_{\bar{i}}^*=(A_i)^*
\end{equation}

Recall finally that this is not the usual complexification of the Dirac
algebra, which is defined with a $1/2$ factor, and hence the lowering 
and
raising operators are given by
\begin{eqnarray}
\sqrt{2}\Gamma_1^{\pm}=\frac{1}{\sqrt{2}}(\Gamma^4\pm i\Gamma^5)
\quad ; \quad
\sqrt{2}\Gamma_2^{\pm}=\frac{1}{\sqrt{2}}(\Gamma^6\pm i\Gamma^7)
\quad ; \quad 
\sqrt{2}\Gamma_3^{\pm}=\frac{1}{\sqrt{2}}(\Gamma^8\pm i\Gamma^9) \quad
\quad
\end{eqnarray}

\end{document}